\documentclass[twocolumn]{aastex631}

\usepackage{xcolor}
\usepackage{verbatim}
\usepackage{amsmath}
\usepackage{hyperref}

\newcommand{\sersic}{S\'{e}rsic}

\begin{document}
\nolinenumbers



\title{Potential-Driven Metal Cycling: JADES Census of Gas-Phase Metallicity for galaxies at $1<z<7$}

\author[0009-0004-7042-4172]{Cheng Jia}\thanks{E-mail:  jc123@mail.ustc.edu.cn}
\affiliation{Department of Astronomy, University of Science and Technology of China, Hefei 230026, China}
\affiliation{School of Astronomy and Space Science, University of Science and Technology of China, Hefei 230026, China}

\author[0000-0003-1588-9394]{Enci Wang}\thanks{Corr-author: ecwang16@ustc.edu.cn}
\affiliation{Department of Astronomy, University of Science and Technology of China, Hefei 230026, China}
\affiliation{School of Astronomy and Space Science, University of Science and Technology of China, Hefei 230026, China}

\author[0009-0000-7307-6362]{Cheqiu Lyu}
\affiliation{Department of Astronomy, University of Science and Technology of China, Hefei 230026, China}
\affiliation{School of Astronomy and Space Science, University of Science and Technology of China, Hefei 230026, China}

\author[0009-0006-7343-8013]{Chengyu Ma}
\affiliation{Department of Astronomy, University of Science and Technology of China, Hefei 230026, China}
\affiliation{School of Astronomy and Space Science, University of Science and Technology of China, Hefei 230026, China}

\author[0000-0002-0846-7591]{Jie Song}
\affiliation{Department of Astronomy, University of Science and Technology of China, Hefei 230026, China}
\affiliation{School of Astronomy and Space Science, University of Science and Technology of China, Hefei 230026, China}

\author[0000-0002-4597-5798]{Yangyao Chen}
\affiliation{Department of Astronomy, University of Science and Technology of China, Hefei 230026, China}
\affiliation{School of Astronomy and Space Science, University of Science and Technology of China, Hefei 230026, China}

\author[0000-0002-3775-0484]{Kai Wang}
\affiliation{Institute for Computational Cosmology, Department of Physics, Durham University, South Road, Durham, DH1 3LE, UK}
\affiliation{Centre for Extragalactic Astronomy, Department of Physics, Durham University, South Road, Durham DH1 3LE, UK}

\author[0009-0008-1319-498X]{Haoran Yu}
\affiliation{Department of Astronomy, University of Science and Technology of China, Hefei 230026, China}
\affiliation{School of Astronomy and Space Science, University of Science and Technology of China, Hefei 230026, China}

\author[0009-0004-5989-6005]{Zeyu Chen}
\affiliation{Department of Astronomy, University of Science and Technology of China, Hefei 230026, China}
\affiliation{School of Astronomy and Space Science, University of Science and Technology of China, Hefei 230026, China}

\author{Jinyang Wang}

\author{Yifan Wang}

\author[0000-0002-7660-2273]{Xu Kong}
\affiliation{Department of Astronomy, University of Science and Technology of China, Hefei 230026, China}
\affiliation{School of Astronomy and Space Science, University of Science and Technology of China, Hefei 230026, China}

\begin{abstract}
The gravitational potential is established as a critical determinant of gas-phase metallicity (12+log(O/H)) in low-redshift galaxies, whereas its influence remains unconfirmed at high redshifts. We investigate the correlation between gas-phase metallicity and effective radius ($R_{\rm e}$) for a sample of galaxies with redshifts ranging from 1 to 7, drawn from JADES (JWST Advanced Deep Extragalactic Survey) Data Release 3. 
We calculate the metallicities using four strong-line methods: ${\rm N2S2H\alpha}$, ${\rm R23}$, ${\rm N2}$, and ${\rm O3N2}$, respectively. 
After taking out the evolution of size, we find that the offsets of mass-size relation ($\Delta \log R_{\rm e}$) are significantly negatively correlated with the offset of mass-metallicity relation ($\Delta \log({\rm O/H})$) for the four metallicity tracers.  
Regardless of the metallicity tracer used, we obtain Spearman rank $p-$values much less than 0.01, rejecting the null hypothesis that the observed correlation is statistically nonsignificant and attributable to random chance.  This is also true for galaxies with $z>3$, with $p-$values less than 0.05 for the four metallicity tracers. 
We for the first time find evidence of size playing a key role in determining gas-phase metallicity towards cosmic dawn, suggesting that the gravitational potential influences their material-exchange processes with the surrounding environment at very early universe.

\end{abstract}

\keywords{}

\section{introduction}\label{sec:intro}


Galaxy evolution is inherently interconnected with environmental processes, rather than occurring in isolation. A well-established body of observational and theoretical studies demonstrates that galaxies dynamically exchange gas and metals with their circumgalactic medium (CGM) through inflows and outflows \citep{Voort_2011, Lilly_2013, Muratov_2017, Mitchell_2018, Wang_2019, Christensen_2018, Veilleux_2020, Thompson_2024, Chen25}. Concurrently, sustained star formation within galaxies drives the continuous enrichment of their metal content. Understanding the coevolution of metal enrichment in galaxies and the broader universe—shaped by these feedback mechanisms—is therefore critical to unraveling the physical drivers of galaxy formation and chemical evolution.

The variations in metallicity within galaxies primarily arise from two factors: exchanges with external matter and internal metal enrichment. The former may be influenced by the accretion of cold gas, feedback from supernovae and active galactic nuclei (AGN), as well as the gravitational potential of the galaxy itself \citep{Bouche_2010, Lilly_2013, Mitchell_2018}. 
The latter is primarily influenced by factors such as star formation history and the initial mass function \citep[IMF;][]{Scannapieco_2005, Wilkins_2022}. 
Although the complex exchange of material between galaxies and their environment remains poorly understood, it is often simplified using gas regulator models \citep{Schaye_2010, Lilly_2013, Peng_2014, Wang_2019, Wang-22a, Wang-22b, Lyu-25}.

As demonstrated by \citet{Wang_2021}, the evolution of gas and metal mass within galaxies can be described by the following relation \citep{Ma_2024}:
\begin{equation}
    \dot{M}_{\rm gas}(t) = \Phi(t) - (1 - R) \Psi(t) - \lambda \Psi(t),
\end{equation}
\begin{equation}
    \dot{M}_{\rm Z}(t) = y \Psi(t) - (1 - R + \lambda) Z(t) \Psi(t) + \Phi(t) Z_0
\end{equation}
where $M_{\rm gas}$ and $M_{\rm Z}$ are the gas mass and metal mass of galaxies, respectively. $\Psi$ is the star formation rate (SFR), $\Phi$ is the inflow rate, $R$ is the fraction of mass returned to the interstellar medium (ISM) from stars, $y$ is the yield and $Z_0$ represents the metallicity of inflowing cool gas. $\lambda$ is the mass-loading factor, which means the ratio between outflow rate and SFR. Assuming that galaxies are stable over relatively long timescales, let $\dot{M}_{\rm gas}(t) = 0$ and $\dot{M}_{\rm Z}(t) = 0$, we can derive:
\begin{equation}
    Z_{\rm gas} = Z_0 + y / (1 - R + \lambda).
\end{equation}
It is evident that the metallicity of galaxies is strongly related to $\lambda$.
In galaxies with deeper gravitational potentials, outflows are suppressed, limiting gas exchange with the environment and leading to metal accumulation; consequently, this model predicts a strong correlation between metallicity and gravitational potential tracers such as stellar mass.

At low redshifts, the mass-metallicity relation (MZR) is well established observationally \citep{Lequeux_1979, Tremonti_2004}, showing that metallicity increases with stellar mass for galaxies with $M_* < 10^{10} M_{\odot}$. Incorporating SFR as a secondary parameter further tightens this relation, leading to the formulation of the fundamental metallicity relation \citep[FMR;][]{Lara_2010, Mannucci_2010, Nakajima_2012, Curti_2020, Ma_2024}. In addition, the effective radius ($R_{\rm e}$)—another key indicator of gravitational potential—also plays a significant role: at fixed stellar mass and SFR, galaxies with smaller $R_{\rm e}$ tend to have higher metallicities \citep{Ellison_2008, Wang-18, DEugenio_2018, Sanchez_2018, Huang_2019, Sanchez_2024a, Sanchez_2024b, Ma_2024, Li-25}. Notably, \cite{Ma_2024} found that $R_{\rm e}$ appears to have a more dominant influence on metallicity than SFR in both observations and simulations, highlighting the crucial role of gravitational potential in regulating galactic chemical enrichment.


For a long time, studies of the MZR have been largely confined to low redshifts, primarily because accurate metallicity measurements require reliable metal emission lines. At high redshifts, these lines are redshifted into the near- and mid-infrared (NIR/MIR), making them difficult to observe from the ground due to atmospheric limitations. Nevertheless, some studies have extended the MZR to $z \sim 3$, revealing that galaxies at these epochs are approximately 0.6 dex more metal-poor than local counterparts at fixed stellar mass \citep{Maiolino_2008, Mannucci_2009, Zahid_2014, Wuyts_2014, Sanders_2021}. The launch of JWST and its NIRSpec instrument has significantly advanced this field by enabling metallicity measurements at higher redshifts. Recent works have leveraged JWST data to explore the chemical properties of early galaxies \citep{Curti_2024, Langeroodi_2023, Nakajima_2023, Chemerynska_2024, Chakraborty_2024, Sanders_2023}. While these efforts have provided a preliminary view of the high-redshift MZR and its evolution, correlations between metallicity and other physical parameters remain uncertain, largely due to limited sample sizes \citep{Nakajima_2023, Curti_2024, Cuestas_2025}.

In this work, we investigate whether galaxy size plays a role in regulating gas-phase metallicity for galaxies at $1 < z < 7$, using data from the JWST Advanced Deep Extragalactic Survey \citep{Eisenstein_2023a, Eisenstein_2023b}.
Since galaxy sizes evolve significantly with redshift and wavelength \citep{Jia_2024}, we account for this by examining the mass–size relation (MSR) at a fixed rest-frame wavelength of $1 \ \mu\mathrm{m}$.
After accounting for the evolution of size, we find a significant negative correlation between metallicity and $R_{\rm e}$ for both samples at $z \sim 1 - 7$ and $z \sim 3 - 7$, suggesting that the gravitational potential affects gas exchange of galaxies in very early universe. 

This paper is organized as follows. In Section \ref{sec:data}, we describe the galaxy sample and the methods used to measure relevant parameters. Section \ref{sec:result} presents our results on the mass-metallicity relation and the correlation between metallicity and galaxy size, as well as the comparison with results of low-redshift and TNG simulations. Finally, in Section \ref{sec:summary}, we summarize our main findings.
In this paper, we assume concordance flat $\mathrm{\Lambda CDM}$ cosmology with $\Omega_{m}=0.308$, $\Omega_{\Lambda}=0.691$, $H_0 = 67.74$ $\mathrm{km \cdot s^{-1} \cdot Mpc^{-1}}$ \citep{Planck_2016}.

\section{data and method}\label{sec:data}

\subsection{JADES data}\label{sec:jadesdata}

In this study, we utilize the publicly available NIRCam and NIRSpec data from the JWST Advanced Deep Extragalactic Survey \citep[JADES; \href{https://doi.org/10.17909/8tdj-8n28}{10.17909/8tdj-8n28};][]{Eisenstein_2023a, Eisenstein_2023b, Rieke_2023, DEugenio_2024}. The survey covers the GOODS-S and GOODS-N ultra-deep fields, providing approximately 125 arcmin$^2$ of near-infrared imaging. As one of the most significant deep-field programs, JADES offers a wealth of information on high-redshift galaxies.


The imaging data used in this study are sourced from the DAWN JWST Archive (DJA) Mosaic Release Version 7, processed with the {\tt grizli} pipeline \citep{Valentino_2023, _zenodo}
To extend the wavelength coverage, we combine imaging from both HST and JWST across ten broad bands—F435W, F606W, F814W, F090W, F115W, F150W, F200W, F277W, F356W, and F444W—covering a wide wavelength range from the optical blue to the near-infrared. All HST and JWST images used in this work are obtained from DJA. 

We adopt the morphological catalogs of the GOODS-S and GOODS-N fields provided by the DJA as our primary sample. To associate galaxy images with their spectra, we cross-match the DJA catalogs with the spectroscopic catalogs from JADES Data Release 3 \citep{DEugenio_2024}. This data is obtained from the Mikulski Archive for Space Telescopes (MAST) at the Space Telescope Science Institute. The specific observations analyzed can be accessed via \dataset[10.17909/8tdj-8n28]{https://doi.org/10.17909/8tdj-8n28}.
Emission line information is drawn from medium-resolution ($R \sim 1000$) NIRSpec spectra, obtained using the F070LP/G140M, F170LP/G235M, and F290LP/G395M filter/grating combinations, which cover wavelength ranges of 0.97–1.84 $\mu$m, 1.66–3.07 $\mu$m, and 2.87–5.10 $\mu$m, respectively. We directly use the emission line measurements from the medium-resolution grating spectra catalogs in JADES DR3 to perform extinction corrections and derive gas-phase metallicities, as detailed in Section \ref{sec.Zmeasure}.


We perform a cross-match using a 0.5 arcsecond radius and exclude any sources with multiple matches within this radius to avoid contamination from nearby objects. Additionally, we remove entries from the spectral catalog flagged with {\tt `DR\_flag'} = {\tt `True'} to ensure data quality. To guarantee reliable redshift measurements, we retain only sources with {\tt `z\_Spec\_flag'} labeled as ‘A’, ‘B’, or ‘C’, corresponding to a redshift confidence level of ‘secure’ or better. This results in a preliminary parent sample of 1,366 galaxies. However, since many of these sources lack reliable emission line measurements, further sample refinement is carried out in the subsequent analysis.

\subsection{Stellar mass measurement}

The stellar masses used in this study are estimated using the CIGALE code \citep{Boquien_2020}. We perform spectral energy distribution (SED) fitting with ten photometric bands (see Section \ref{sec:jadesdata}), combining HST and JWST data. The fitting process adopts a delayed exponential star formation history, \cite{Bruzual_Charlot_2003} stellar population synthesis models, the \cite{Calzetti_2000} dust attenuation law, and the \cite{Inoue_2011} nebular emission line models.
To ensure a more reliable sample, we apply the following selection criteria:
\begin{enumerate}
    \item $S/N_{\rm det} > 10$, where $S/N_{\rm det}$ is the signal-to-noise ratio in the detection image.

    \item $\chi^2/N_{\rm filter} < 8$ and $N_{\rm filter} > 6$ in photometric redshift estimation.

    \item ${\rm mag_{F444W} < 28.5}$.

    \item $S/N > 3$ for F115W, F150W, F200W, F277W, F356W, and F444W.
\end{enumerate}
Further details of the physical parameter estimation process will be presented in an upcoming work (Song et al., in prep.).


\subsection{Size measurement}

Galaxy sizes can vary significantly with wavelength, reflecting the spatial distribution of different stellar populations within galaxies \citep{Jia_2024}. To ensure consistency, we define galaxy sizes at a fixed rest-frame wavelength whenever possible. However, achieving this across a range of redshifts requires using different observed bands, each with its own point spread function (PSF). To address this, we construct PSFs for each band using {\tt photutils.psf.EPSFBuilder} and match them to the F444W band using {\tt photutils.psf.create\_matching\_kernel}. This process ensures that all images are homogenized to a common resolution.

We measure the size of the galaxies using {\sc Galfit} \citep{Peng_2002, Peng_2010}, fitting a \sersic\ profile model:
\begin{equation}
    \Sigma (r) = \Sigma_{\mathrm{e}} \exp{\left[ -\kappa \left(\left(\frac{r}{R_{\mathrm{e}}}\right)^{1/n} - 1\right) \right]}, 
\end{equation}
where $n$ is the \sersic\ index, $\kappa$ is a function of \sersic\ index, and $R_{\mathrm{e}}$ represents the 2D half-light radius measured along the semi-major axis. We simultaneously determine the center, magnitude, effective radius, \sersic\ index, axis ratio, and position angle while fitting. The initial values used for fitting were obtained from the morphological catalogs provided by DJA. The error map used for fitting was obtained from the inverse of the square root of the weight map supplied by DJA, and we also employ DJA's segmentation map to mask the influence of other sources.


We adopt the effective radius at a rest-frame wavelength of $1\ {\rm \mu m}$ as our measure of galaxy size, as this wavelength predominantly traces older, more stable stellar populations and provides a robust proxy for the underlying mass distribution and gravitational potential. For each galaxy, we estimate the $1\ {\rm \mu m}$ size via linear interpolation between observed bands bracketing the rest-frame wavelength. If only one side yields a successful measurement, we adopt that value; if both fail, the size is marked as undetermined. For galaxies at $z > 3.5$, where all observed bands correspond to rest-frame wavelengths shorter than $1\ {\rm \mu m}$, we directly use the size measured in the F444W band—the longest available wavelength—as an approximation. Of the 1,366 galaxies in our sample, we successfully derive $1\ {\rm \mu m}$ sizes for 1,123.

\subsection{Metallicity measurement}\label{sec.Zmeasure}

To improve the robustness of our results, we adopt multiple methods for calculating the gas-phase metallicity of galaxies. We correct the extinction of emission line fluxes using the extinction law from \cite{ODonnell_1994}, adopting an intrinsic Balmer decrement of ${\rm H\alpha/H\beta} = 2.87$. 
For metallicity estimation, we utilize several line ratio diagnostics, including ${\rm N2S2H\alpha}$, ${\rm N2}$, ${\rm R23}$, and ${\rm O3N2}$ \citep{Dopita_2016, Steidel_2014, Kobulnicky_2004, Kewley_2002}. We note that all galaxies with valid emission line measurements in the JADES DR3 catalog have signal-to-noise ratios ($S/N$) greater than 5. Thus, no additional $S/N$ cuts are applied.

We utilize the metallicity calibration from \cite{Dopita_2016}:
\begin{equation}
    12 + {\rm \log (O/H)} = 8.77 + {\rm N2S2H\alpha} + 0.45 ({\rm N2S2H\alpha} + 0.3)^5
\end{equation}
where ${\rm N2S2H\alpha} = \log ({\rm [NII] 6584 / [SII] 6717,6731}) + 0.264 \log ({\rm [NII] 6584 / [H\alpha]})$. This indicator is regarded as highly effective for estimating the metallicity of HII regions due to its insensitivity to reddening \citep{Wang_2021, Easeman_2023, Ma_2024}.

Additionally, we employ the formulas provided by \cite{Steidel_2014}:
\begin{equation}
    12 + {\rm \log (O/H)} = 8.62 + 0.36 {\rm N2},
\end{equation}
\begin{equation}
    12 + {\rm \log (O/H)} = 8.66 - 0.28 {\rm O3N2}
\end{equation}
where $N2 = \log ({\rm [NII] 6584 / [H\alpha]})$ and ${\rm O3N2} = \log ({\rm [OIII] 5007 / [H\beta]}) - \log ({\rm [NII] 6584 / [H\alpha]})$. These indicators are widely used for estimating metallicity and are valid for the ranges $-1.7 < N2 < -0.3$ and $-0.4 < {\rm O3N2} < 2.1$.

For ${\rm R23}$ 
$(=\log [({\rm [OII]3727 + [OIII]5007) / [H\beta]}])$, we use the following calibrations for the upper and lower branches, respectively \citep{Kobulnicky_2004}:
\begin{equation}
\begin{aligned}
    12 + {\rm \log (O/H)}_{\rm upper} & =  9.72 - 0.777x - 0.951x^2 \\
    & - 0.072x^3 - 0.811x^4 \\
    & -\log(q)(0.0737 - 0.0713x \\
    & - 0.141x^2 + 0.0373x^3 - 0.058x^4), \\
    12 + {\rm \log (O/H)}_{\rm lower} & = 9.40 + 4.65x - 3.17x^2 \\
    &- \log(q)(0.272 + 0.547x - 0.513x^2)
\end{aligned}
\end{equation}
where $x = {\rm R23}$, $q$ is ionization parameter given from \cite{Kobulnicky_2004}.
To address the degeneracy of the ${\rm R23}$ method, we use a threshold of $N2 > -0.8$ to distinguish between the upper and lower branches \citep{Kewley_2002}.

\begin{figure}[ht]
    \centering
    \includegraphics[scale=0.4]{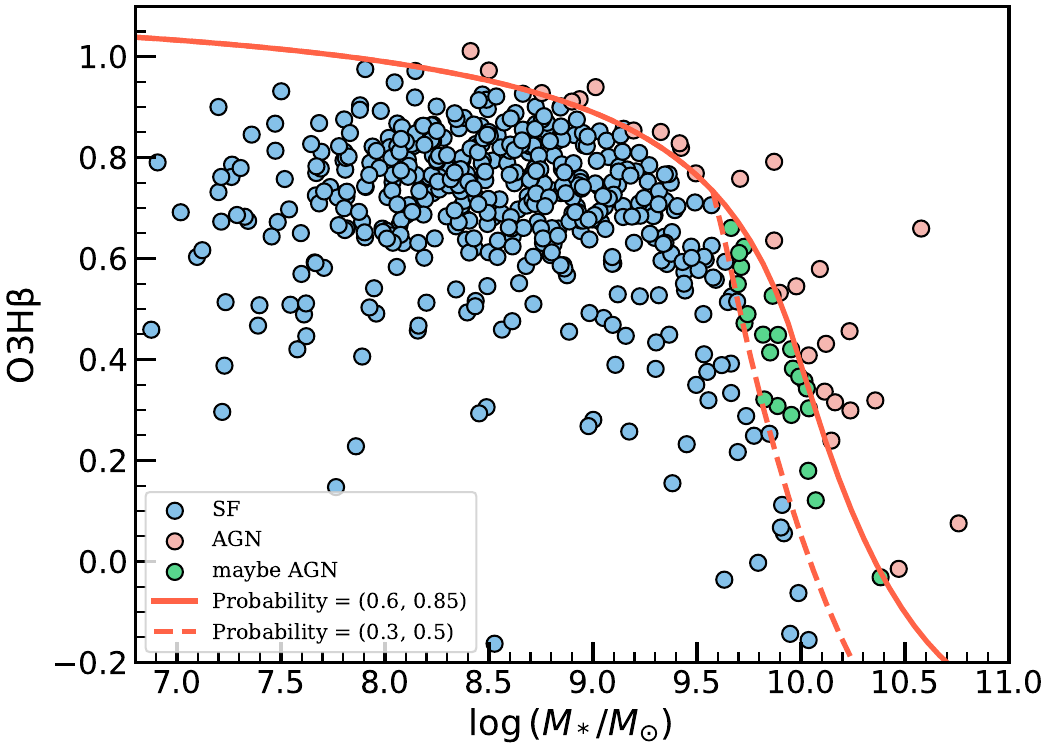}
    \caption{The mass-extinction diagram for galaxies with ${\rm O3H\beta}$ measurements. Sky-blue points represent sources classified as star-forming galaxies, while pink points indicate sources identified as AGNs. Green points correspond to potential AGN candidates. The solid and dashed red lines show the high- and low-probability AGN classification thresholds from \citet{Juneau_2014}, respectively. In this work, we adopt the solid red line as our selection criterion.}
    \label{Fig.MEx}
\end{figure}

To account for potential AGN contamination in our sample, we apply the mass–excitation (MEx) diagnostic diagram from \cite{Juneau_2014}. As shown in Figure \ref{Fig.MEx}, the solid and dashed red lines denote thresholds for high and low AGN probability, respectively. We adopt the solid line to exclude likely AGN hosts (pink points), while retaining galaxies with lower AGN likelihood (sky-blue and green points). In addition, we remove sources lacking a measurement of ${\rm O3H\beta}$ $(=\log({\rm [OIII]5007 / [H\beta]}))$ to ensure the purity of the sample.


Since different metallicity calibration methods require distinct sets of emission lines, our final selection combines the criteria outlined in Section \ref{sec:jadesdata} with additional constraints: each galaxy must have valid measurements of stellar mass, effective radius, and relevant emission lines, and must not be classified as an AGN. To further ensure sample purity, we cross-match our selection with the X-ray AGN catalog from \cite{Luo_2016}, identifying and removing one additional AGN not flagged by the MEx diagram. The final sample sizes for each calibration method are as follows: 75 galaxies for ${\rm N2S2H\alpha}$, 97 for ${\rm N2}$, 77 for ${\rm R23}$, and 94 for ${\rm O3N2}$, as illustrated in Appendix \ref{App.sample} and Figure \ref{Fig.sample}. All galaxies in the final sample have stellar masses exceeding $10^8\ M_{\odot}$. 
The information about the galaxies used in this work is presented in Appendix \ref{App.table}, and more information related to emission lines can be accessed via \href{https://drive.google.com/file/d/1_X__k1Adm_l-1Mz2Y-oq7b7gW-Gtk8TN/view?usp=share_link}{this link}\footnote{\href{https://drive.google.com/file/d/1_X__k1Adm_l-1Mz2Y-oq7b7gW-Gtk8TN/view?usp=share_link}{https://drive.google.com/file/d/1\_X\_\_k1Adm\_l-1Mz2Y-oq7b7gW-Gtk8TN/view?usp=share\_link}}.

\section{results and discussions}\label{sec:result}

\subsection{Mass-metallicity relation}

\begin{figure*}[ht]
    \centering
    \begin{minipage}[c]{1\textwidth}
        \includegraphics[width=0.5\textwidth]{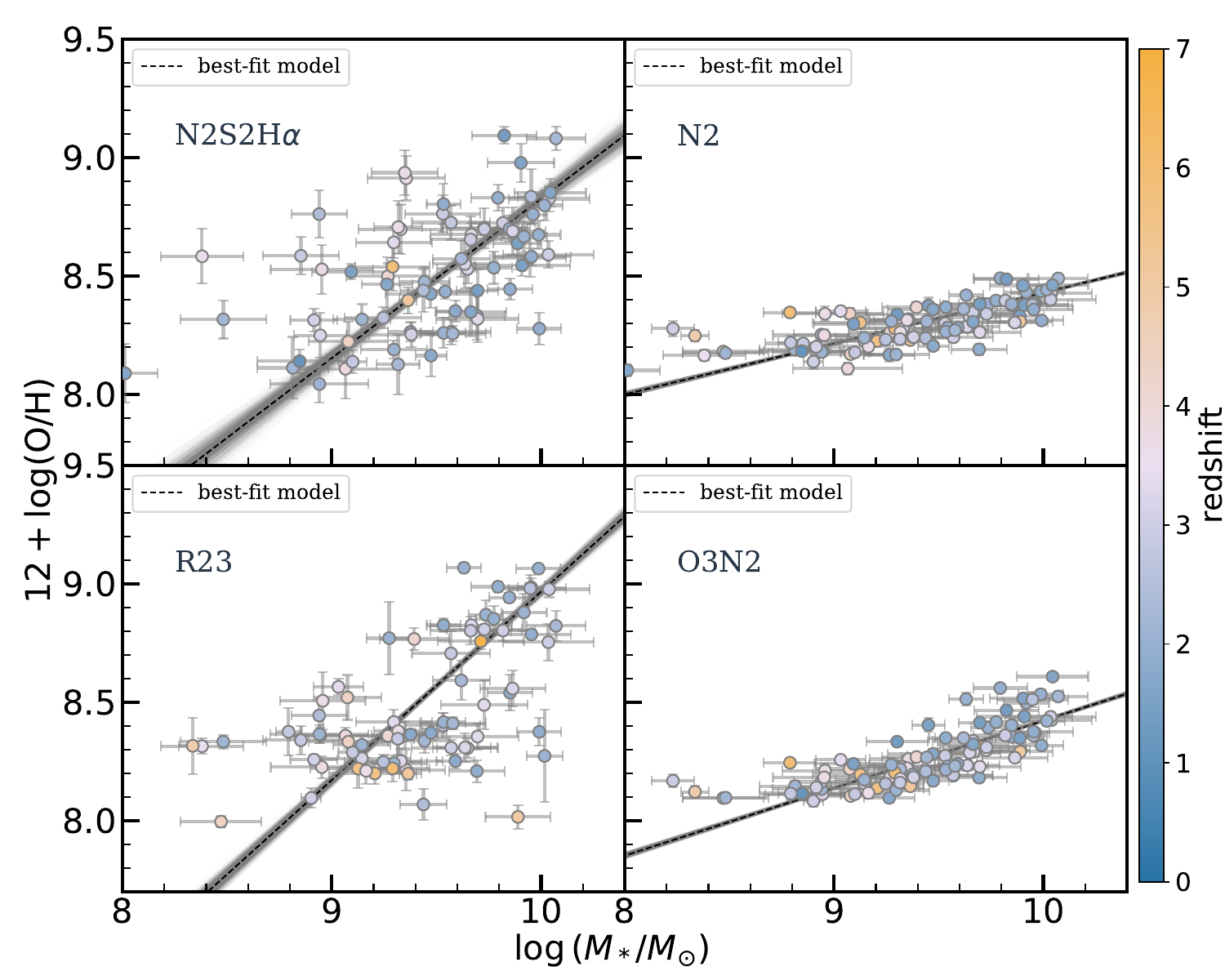}
        \includegraphics[width=0.5\textwidth]{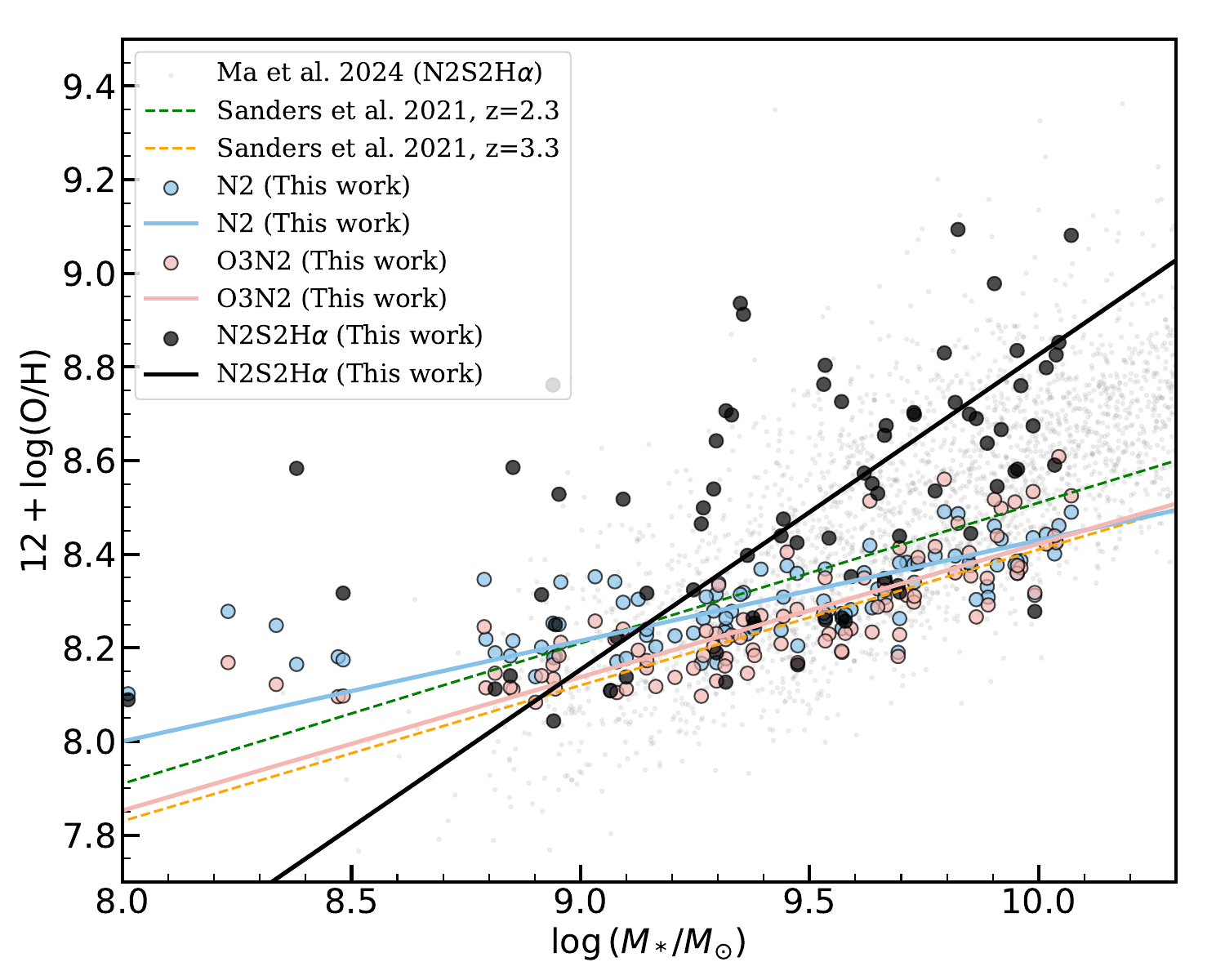}
    \end{minipage}
    \caption{The MZR measured using the four metallicity indicators and a comparison with others. Solid dots in the left panel represent measurements from individual sources, while the black dashed lines indicate the best-fit results. The gray shade around best-fit model represents the error. Data points are color-coded with the redshift of galaxies.
    In the right panel, the solid points in various colors and the corresponding solid lines represent the individual source measurements and their best-fit models for each method, just the same as the left panel. The dashed green and orange lines indicate the MZR provided by \cite{Sanders_2021} at redshifts of 2.3 and 3.3. The translucent gray points represent the results with MaNGA using ${\rm N2S2H\alpha}$ \citep{Ma_2024}.}
    \label{Fig.combined}
\end{figure*}


Different metallicity calibration methods are known to exhibit systematic offsets \citep[e.g.,][]{Kewley_2008, Sanders_2021}, which limits the significance of direct comparisons across methods. These four metallicity calibration methods serve to cross-validate the reliability of our results. In addition, for each metallicity tracer, a key advantage of this calibration approach is its ability to utilize a consistent set of line ratios across galaxies spanning a wide redshift range. 

Figure 
\ref{Fig.combined} presents the MZR with the color-coding of redshifts derived from the four different metallicity calibrations. We perform a linear fit to characterize the MZR across the full redshift range of our sample using the Markov chain Monte Carlo (MCMC) method, incorporating a 2D Gaussian likelihood that accounts for the uncertainties in both $\log (M_*/M_{\odot})$ and $\log({\rm O/H})$. 
As shown, we find no significant evidence of metallicity evolution with redshift for any of the metallicity indicators employed.
The MZRs based on ${\rm N2}$ and ${\rm O3N2}$ exhibit notably tight correlations, with ${\rm O3N2}$ yielding slightly higher metallicities than ${\rm N2}$ \citep{Bian_2018}. In contrast, the MZRs derived from ${\rm N2S2H\alpha}$ and ${\rm R23}$ are significantly steeper. At a stellar mass of approximately $10^{8.5}\ M_{\odot}$, all four indicators converge around $12 + \log({\rm O/H}) \sim 8.2$. However, at $\sim10^{10}\ M_{\odot}$, the ${\rm N2}$ and ${\rm O3N2}$ indicators yield metallicities around $12 + \log({\rm O/H}) \sim 8.5$, while some galaxies measured with ${\rm N2S2H\alpha}$ and ${\rm R23}$ reach up to $12 + \log({\rm O/H}) \sim 9.0$.

The ${\rm R23}$ calibration is known to suffer from the degeneracy between its upper and lower branches, making measurements around $12 + \log({\rm O/H}) \sim 8.5$ particularly uncertain \citep{Kobulnicky_2004}. In comparison, ${\rm N2S2H\alpha}$ is regarded as a more robust tracer for HII region metallicities, showing only a systematic offset from $T_{\rm e}$-based estimates that is largely independent of ionization parameter \citep{Easeman_2023}. Despite this, the MZR obtained from ${\rm N2S2H\alpha}$ shows a steeper slope and larger scatter at $M_\ast < 10^{10}\ M_{\odot}$ compared to those based on ${\rm N2}$ and ${\rm O3N2}$ \citep{Ma_2024}.

Figure 
\ref{Fig.combined} presents a comparison between our results and those from previous studies. Solid points in different colors represent our measurements based on various metallicity indicators, with corresponding solid lines showing the best-fit models. The median redshifts for the samples used in the ${\rm N2S2H\alpha}$, ${\rm N2}$, and ${\rm O3N2}$ measurements are 2.27, 2.62, and 2.60, respectively. The green and orange lines indicate the MZR from \citet{Sanders_2021} at redshifts 2.3 and 3.3. Our results based on ${\rm N2}$ and ${\rm O3N2}$ are in good agreement with those of \citet{Sanders_2021}, who used four emission lines (${\rm [OII]}$, ${\rm [H\beta]}$, ${\rm [OIII]}$, and ${\rm [NeIII]}$) to minimize the $\chi^2$ in their metallicity calibration. In contrast, the ${\rm N2S2H\alpha}$ measurements show a notable deviation from the other indicators. Interestingly, the MZR derived from ${\rm N2S2H\alpha}$ aligns closely with local universe results from MaNGA \citep{Ma_2024}, which also used the ${\rm N2S2H\alpha}$ calibration. This suggests that when metallicity is estimated using ${\rm N2S2H\alpha}$, there may be little to no apparent evolution in metallicity with redshift.

\subsection{Correlation between size and metalllicity}\label{sec.dZdR}

\begin{figure*}[ht]
    \hspace*{\fill}
    \includegraphics[width=0.9\textwidth]{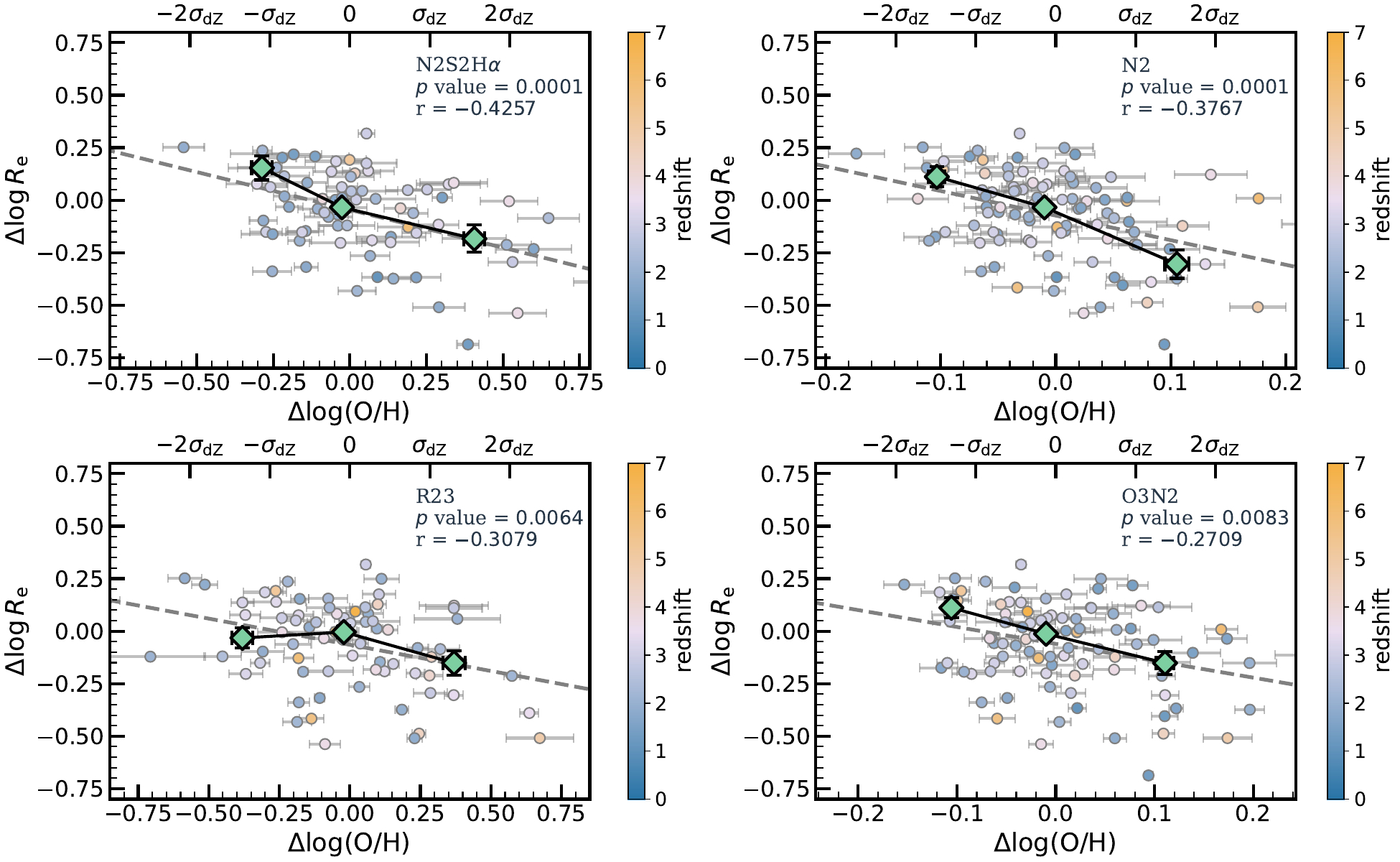}
    \hspace*{\fill}
    \caption{Correlation between $\Delta {\rm \log (O/H)}$ and $\Delta \log R_{\rm e}$. Solid circular points represent individual galaxies, color-coded by redshift. Green diamonds indicate the median values within three bins: $(-3\sigma_{\rm dZ}, -\sigma_{\rm dZ})$, $(-\sigma_{\rm dZ}, \sigma_{\rm dZ})$, and $(\sigma_{\rm dZ}, 3\sigma_{\rm dZ})$. The Spearman rank $p$-values are displayed in the upper left corner of each panel.}
    \label{Fig.dZdR}
\end{figure*}

At low-reshift universe, incorporating the radius $R_{\rm e}$ as a third parameter in the MZR significantly tightens the relationship, highlighting the connection between gravitational potential $(=M_*/R_{\rm e})$ and gas-phase metallicity \citep{Ellison_2008, DEugenio_2018, Huang_2019, Ma_2024}.  For high-redshift galaxies, the limited sample size makes direct verification through fitting challenging. As a result, we investigate this by examining the correlation between a galaxy's deviation from the MZR and its deviation from the MSR.

We define $\Delta {\rm \log (O/H)}$ as the deviation of galaxies from the fitted MZR, the best-fit lines in Figure 
\ref{Fig.combined} This parameter indicates whether a galaxy has higher or lower metallicity than the average at a fixed stellar mass. To account for the strong redshift dependence of galaxy size, we define $\Delta \log R_{\rm e}$ as the deviation from the mass-size relation at the correpsonding redshift. We then calculate $\Delta \log R_{\rm e}$ by adopting a size evolution model that includes both stellar mass and redshift—an updated version of \cite{Jia_2024} incorporating a larger JWST dataset: 
\begin{equation}
    \Delta \log R_{\rm e} = \log R_{\rm e} - \log R_{\rm model},
\end{equation}
\begin{equation}
    \log R_{\rm model} = \alpha \log (M_*/M_{\odot}) + \beta \log (1+z) + k
\end{equation}
where $\alpha=0.162, \beta=-0.614, k=-0.964$ (Song et al. in prep), and $R_{\rm e}$ is the effective radius at $1\ {\rm \mu m}$. 
This quantity reflects how a galaxy’s gravitational potential deviates from the average at its corresponding redshift. Galaxies with smaller $\Delta \log R_{\rm e}$ values are more compact than typical galaxies of the same mass and redshift. If gravitational potential plays a role in regulating metallicity, we would expect a negative correlation between $\Delta \log ({\rm O/H})$ and $\Delta \log R_{\rm e}$, implying that more compact galaxies tend to have higher metallicities.

\begin{figure*}[ht]
    \hspace*{\fill}
    \includegraphics[width=\textwidth]{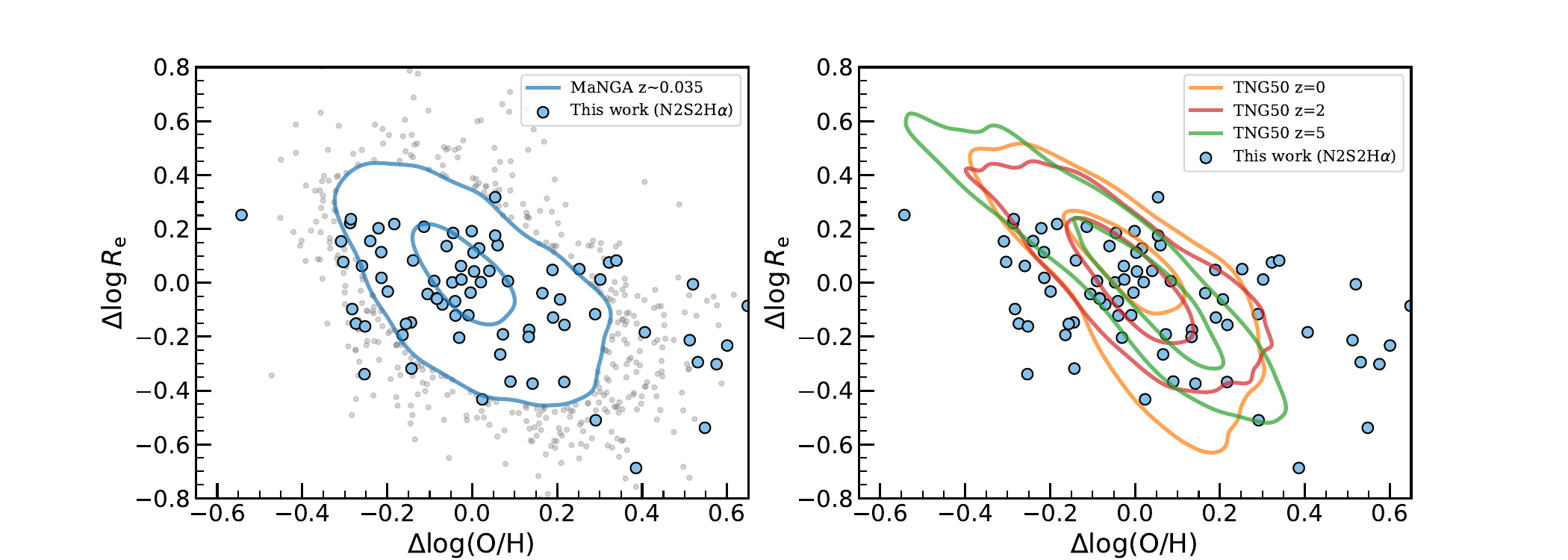}
    \hspace*{\fill}
    \caption{
    Similar to Figure \ref{Fig.dZdR}, but showing a comparison between our results and those from local observations and simulations. Blue dots represent our measurements using the ${\rm N2S2H\alpha}$ indicator. In the left panel, the blue contour and gray points show results from MaNGA \citep{Ma_2024}. In the right panel, colored contours represent TNG simulation results across different redshift slices.}
    \label{Fig.compare_dzdr}
\end{figure*}

Figure \ref{Fig.dZdR} shows the correlations between $\Delta {\rm \log (O/H)}$ and $\Delta \log R_{\rm e}$ for the four metallicity indicators. The solid circular points and their colors are the same as in Figure 
\ref{Fig.combined}. The dashed line represents the results of a least-squares linear fitting. The green diamonds indicate the median values of $\Delta {\rm \log (O/H)}$ and $\Delta \log R_{\rm e}$ for the three bins: $(-3\sigma_{\rm dZ}, -\sigma_{\rm dZ})$, $(-\sigma_{\rm dZ}, \sigma_{\rm dZ})$, and $(\sigma_{\rm dZ}, 3\sigma_{\rm dZ})$, where $\sigma_{\rm dZ}$ is the scatter of $\Delta {\rm \log (O/H)}$ for each metallicity indicator. The error bars are calculated using $\sigma/\sqrt{N}$, where $N$ is the number of galaxies in each bin. 

Despite systematic offsets among different metallicity calibration methods, we consistently find a significant negative correlation between metallicity and galaxy size across all methods, with Spearman rank $p$-values below 0.01. This suggests that the likelihood of observing such a correlation by chance—if no intrinsic relation existed—is less than 1\%. Notably, the ${\rm N2S2H\alpha}$ and ${\rm N2}$ indicators yield particularly strong correlations, with $p$-values around 0.0001, indicating a robust negative correlation between metallicity and galaxy size. 
This strongly supports the conclusion that, within this redshift range, a galaxy's metallicity is significantly linked to its gravitational potential.

We note that fitting the MZR using galaxies spanning a wide redshift range may introduce potential biases, even though our sample shows no significant metallicity evolution with redshift. Specifically, this approach could underestimate the slope of the $\Delta {\rm \log (O/H)}$–$\Delta \log R_{\rm e}$ relation, as any unaccounted-for redshift evolution in metallicity may contribute additional scatter. However, since we have corrected for the evolution of galaxy size with redshift, such effects are unlikely to produce artificial correlations. As a result, our conclusions remain robust.
To further mitigate the influence of low- to intermediate-redshift galaxies dominating the signal, we perform the same analysis on galaxies at $z > 3$ separately. Details of this high-redshift analysis are provided in Appendix \ref{App.high-z}. We find that even with a smaller sample size, the negative correlation between $\Delta {\rm \log (O/H)}$ and $\Delta \log R_{\rm e}$ persists for galaxies of $3<z<7$ with $p-$values less than 0.05 for the four metallicity indicators, reinforcing the validity of our findings.

\subsection{Comparison with low-redshift and simulations}

We compare our results with local observation MaNGA \citep[Mapping Nearby Galaxies at Apache Point Observatory;][]{Bundy-15} and TNG50 (the
Illustris The Next Generation with highest resolution) simulation \citep{Nelson-18, Pillepich-18}, taken from \cite{Ma_2024}. The result is shown in Figure \ref{Fig.compare_dzdr}.  

We define $\Delta {\rm \log (O/H)}$ and $\Delta \log R_{\rm e}$ for MaNGA galaxies in a similar way, by  fitting the MSR and MZR with a fourth-order polynomial \citep{Ma_2024}. 
A significant negative correlation between $\Delta {\rm \log (O/H)}$ and $\Delta \log R_{\rm e}$ is clearly observed for MaNGA galaxies, with a slope of approximately $-0.8$, consistent with the findings of \cite{Ma_2024}. Our high-redshift results broadly agree with those from local galaxies, though the slope appears slightly shallower. This difference may arise from the additional scatter introduced by the potential evolution of metallicity with redshift (see Section \ref{sec.dZdR}), and/or the intrinsic difference between low- and high-redshift. 

We further compare our results with predictions from the TNG50 simulations, shown in the right panel of Figure \ref{Fig.compare_dzdr}. For TNG50, we compute $\Delta {\rm \log (O/H)}$ and $\Delta \log R_{\rm e}$ based on the redshift-dependent linear MSR and MZR. Notably, the slopes of the $\Delta {\rm \log (O/H)}$–$\Delta \log R_{\rm e}$ relation in the simulations converge across different redshifts, again consistent with \cite{Ma_2024}. This agreement reinforces the robustness of our method for identifying the negative correlation between metallicity and size.

\begin{figure}[ht]
    \centering
    \includegraphics[scale=0.45]{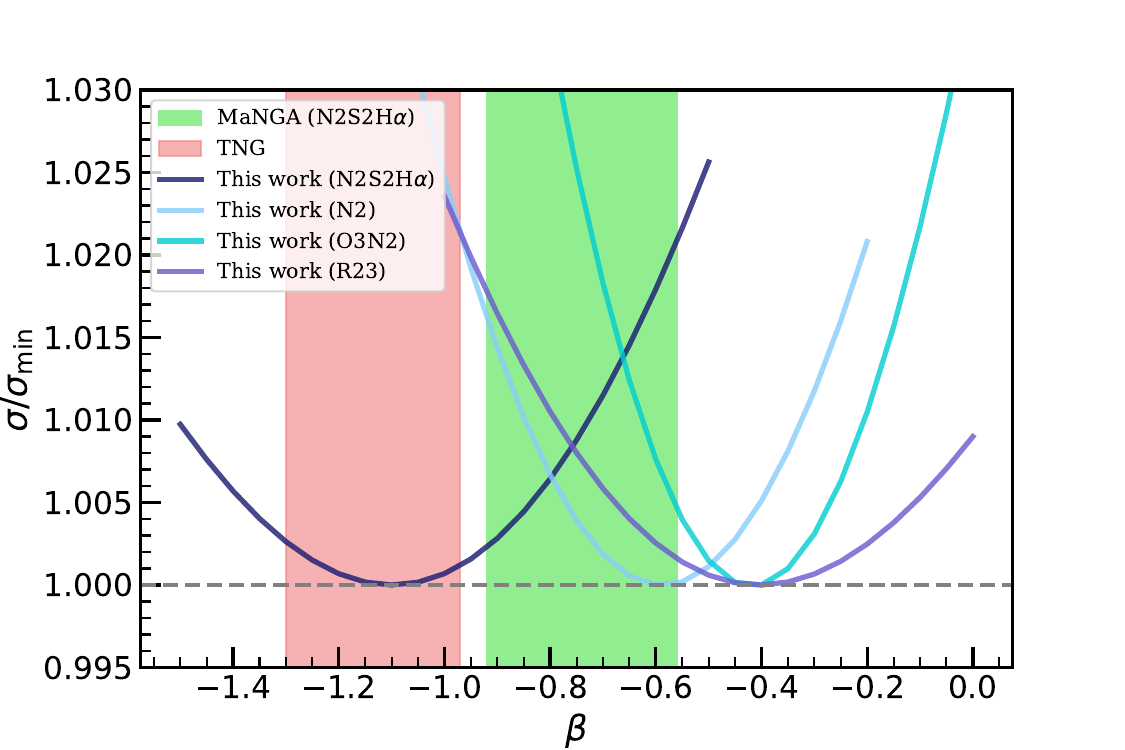}
    \caption{
    The relative scatter in residual $\log {\rm (O/H)}$, denoted as $\sigma/\sigma_{\rm min}$, as a function of $\beta$. Solid lines in different colors show results obtained using various metallicity indicators. The pink shaded region represents the best-fit $\beta$ range for redshifts $z = 0$–5 from the TNG50 simulations.  The green shaded region marks the 1$\sigma$ range around the best-fit $\beta$ in MaNGA using the ${\rm N2S2H\alpha}$ indicator. Data for both TNG50 and MaNGA are taken from \cite{Ma_2024}.
    }
    \label{Fig.beta}
\end{figure}

Having confirmed the correlation between size and metallicity, we proceed to quantitatively characterize this relationship in order to explore its underlying physical drivers. Given the lack of evidence that SFR significantly influences metallicity at high redshift \citep{Cuestas_2025}, we perform a least-squares fit to a modified MZR of the following form:
\begin{equation}
    12 + \log ({\rm O/H}) = k x + b,
\end{equation}
\begin{equation}
    x = \log (M_*/M_{\odot}) + \beta \Delta \log R_{\rm e}.
\end{equation}
We explore a range of $\beta$ values to determine the best-fit relation with minimizing the corresponding residual scatter of metallicity. The results are shown in Figure \ref{Fig.beta}. For comparison, we also show the MaNGA and TNG50 results from \cite{Ma_2024}, which take out the effect of SFR. Using ${\rm N2}$, ${\rm O3N2}$, and ${\rm R23}$ as metallicity indicators, we find that the best-fit $\beta$ values range from approximately $-0.6$ to $-0.4$, consistent with previous results at low redshift \citep{DEugenio_2018, Sanchez_2024b}. Interestingly, when using ${\rm N2S2H\alpha}$, the best-fit $\beta$ is around $-1.1$, closely aligning with trends seen in the TNG simulations. We conclude that $\beta$ lies in the range of $-0.4$ to $-1.1$, broadly consistent with previous studies. This supports the idea that gravitational potential plays a key role in regulating gas-phase metallicity even in the early universe.  This further highlights that achieving convergence among metallicity calibrators is crucial for unveiling the underlying mechanisms driving this correlation.


\subsection{Potential Impact of specific SFR Gradients}\label{sec:discussion}

More compact galaxies may undergo bursty star formation in their centers, which requires significant gas accretion. As a result, these galaxies are expected to have lower metallicities, as found in the work of \cite{Langeroodi_2023_1}. This effect could weaken or even reverse the correlation between galaxy size and metallicity of our work. On the other hand, if more compact galaxies experience bursty star formation in their centers, they would exhibit more pronounced sSFR gradients, complicating the investigation of the relationship between galaxy size and metallicity. To address this, we divide our sample into two subsamples based on $\Delta \log R_{\rm e}$. We use {\tt Bagpipes} \citep{Carnall_2018} for spatially resolved SED fitting of these galaxies to examine their specific SFR (sSFR) radial gradients. We find no significant differences in sSFR gradients between the two subsamples, suggesting that the difference in star formation regions might do not impact our conclusions. However, given the limited sample size, more extensive spectroscopic observations will be needed in the future to draw more robust conclusions.

\section{summary}\label{sec:summary}

We combine imaging and spectroscopic data from the JADES field to investigate the metallicity and morphological properties of galaxies at redshifts of $1<z<7$. Our analysis reveals a significant negative correlation between galaxy size and gas-phase metallicity across this redshift range, suggesting that gravitational potential has played a crucial role in regulating the metal-enrichment processes of galaxies in the early universe. Our main findings are summarized as follows:

\begin{enumerate}
    \item 
    We derive the metallicities of sample galaxies using four calibration methods: ${\rm N2S2H\alpha}$, ${\rm R23}$, ${\rm N2}$, and ${\rm O3N2}$. These methods yield metallicity estimates with notable quantitative discrepancies. The mass–metallicity relations obtained from ${\rm N2}$ and ${\rm O3N2}$ are consistent with previous studies at comparable redshifts. In contrast, the MZR based on ${\rm N2S2H\alpha}$ shows no clear evidence of evolution relative to local galaxy samples.

    \item 
    We investigate the relationship between the deviation from the mass–size relation ($\Delta \log R_{\rm e}$) and the deviation from the mass–metallicity relation ($\Delta {\rm \log (O/H)}$) of galaxies. Despite variations in metallicity estimates across different calibration methods, all four approaches consistently reveal a significant negative correlation between $\Delta {\rm \log (O/H)}$ and $\Delta \log R_{\rm e}$ with Spearman rank $p-$values less than 0.01. This indicates that, at fixed stellar mass and redshift, more compact galaxies tend to exhibit higher metallicities. This is also true for galaxies at $3<z<7$. 
    
    \item We compare our results with those from the local universe (MaNGA) and the TNG50 simulations, finding broad consistency across all datasets. Furthermore, our quantitative analysis reveals that metallicity is closely linked to the parameter $\log (M_*/M_{\odot}) + \beta \Delta \log R_{\rm e}$, with $\beta$ ranging from $-0.4$ to $-1.1$. This strongly supports the idea that gravitational potential plays a key role in regulating metallicity, likely by confining metal-rich outflows.  
    
\end{enumerate}


Our study offers new insights into the relationship between galaxy size and metallicity towards the cosmic dawn, providing preliminary confirmation of their correlation. To refine these findings and better understand how metallicity relates to other physical properties—especially at even higher redshifts—future research will require larger, higher-quality spectroscopic samples. We anticipate that upcoming JWST observations will significantly advance our understanding of the processes governing galaxy metallicity enrichment. This study also highlights the necessity of consistent metallicity calibrations to effectively address such challenges.

\section{acknowledgements}\label{sec:acknowledgements}
Data used in this work are obtained from the Mikulski Archive for Space Telescopes (MAST). The spectrums are from the JADES Data Release 3 archive, which is available at \href{https://jades-survey.github.io/scientists/data.html}{https://jades-survey.github.io/scientists/data.html}. The images products are retrieved from DJA.

EW thanks support of the National Science Foundation of China (Nos. 12473008) and the Start-up Fund of the University of Science and Technology of China(No. KY2030000200). The authors gratefully acknowledge the support of Cyrus Chun Ying Tang Foundations.

\appendix

\section{Sample distribution}\label{App.sample}

\begin{figure*}[ht]
    \hspace*{\fill}
    \includegraphics[width=0.85\textwidth]{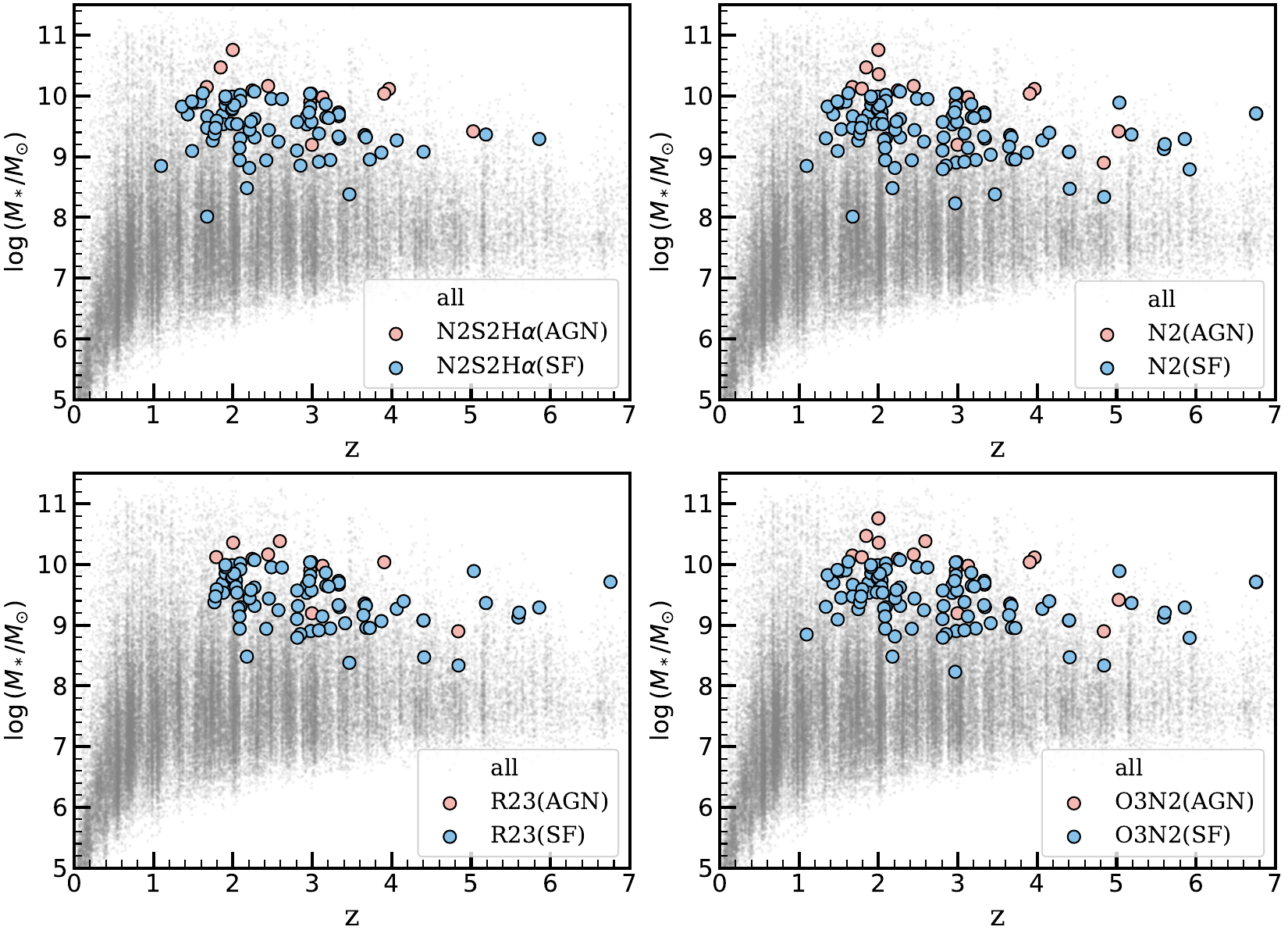}
    \hspace*{\fill}
    \caption{The redshift–stellar mass diagram of our sample. Sky-blue and pink points represent star-forming galaxies and AGNs, respectively, from the final samples used for different metallicity measurement methods. Gray dots indicate all galaxies in the JADES field with available stellar mass and photometric redshift measurements.}
    \label{Fig.sample}
\end{figure*}

\begin{figure}[ht]
    \centering
    \includegraphics[scale=0.45]{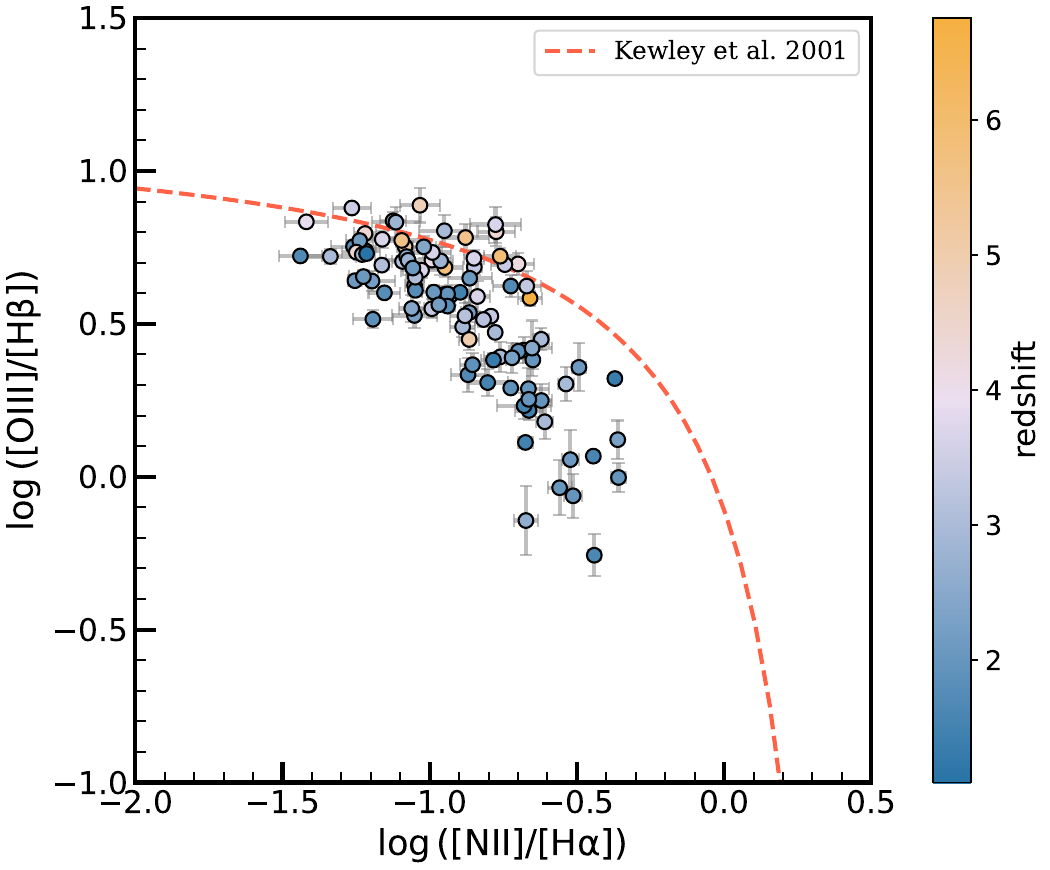}
    \caption{The BPT diagram of our sample. This includes all the galaxies used in our four methods.}
    \label{Fig.BPT}
\end{figure}

The Figure \ref{Fig.sample} illustrates the distribution of the final sample described in Section \ref{sec:data}. The four panels present the samples used for the indicators ${\rm N2S2H\alpha}$, ${\rm N2}$, ${\rm R23}$, and ${\rm O3N2}$. All samples have stellar masses greater than $10^8 M_{\odot}$. The redshift distribution ranges from 1 to 7, with the median redshifts for the four indicators being: 2.27 for ${\rm N2S2H\alpha}$, 2.62 for ${\rm N2}$, 2.93 for ${\rm R23}$, and 2.60 for ${\rm O3N2}$. The BPT diagram of our sample is shown in Figure \ref{Fig.BPT}. The distribution of our sample is similar with that of star-forming galaxies at high redshift \citep{Steidel_2014, Sanders_2023_BPT}, indicating that our sample is not contaminated by AGNs.

\section{Test for galaxies with redshift greater than 3}\label{App.high-z}


To minimize the influence of relatively low-redshift galaxies on our results, we conduct a separate analysis focusing solely on galaxies at redshifts $z > 3$. The MSR and MZR relations used for this analysis are consistent with those described in Section \ref{sec.dZdR}. Despite the limited sample sizes—21 for ${\rm N2S2H\alpha}$, 34 for ${\rm N2}$, 32 for ${\rm R23}$, and 32 for ${\rm O3N2}$—a significant negative correlation between $\Delta {\rm \log (O/H)}$ and $\Delta \log R_{\rm e}$ remains evident. All four methods yield $p$-values below 0.05, indicating that, even within the redshift range $z = 3$–$7$, galaxy size remains a significant factor influencing gas-phase metallicity, likely through its connection to gravitational potential. 

\begin{figure*}[ht]
    \hspace*{\fill}
    \includegraphics[width=0.9\textwidth]{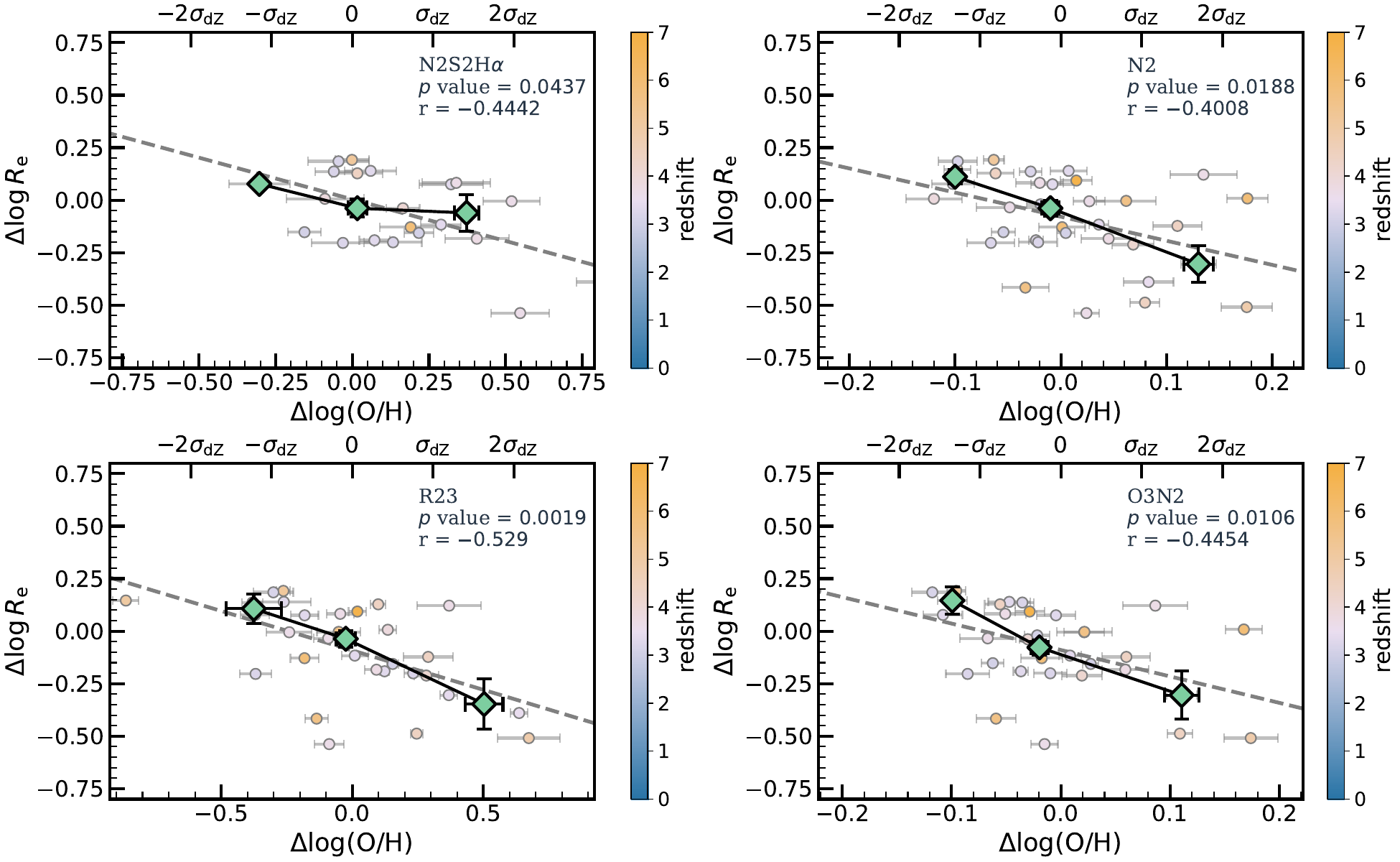}
    \hspace*{\fill}
    \caption{Similar to Figure \ref{Fig.dZdR}, but for galaxies $z > 3$.}
    \label{Fig.highz}
\end{figure*}


\section{List of properties of our sample}\label{App.table}

\begin{table*}[h]
\movetabledown=2.0in
\begin{rotatetable*}
\begin{center}
\begin{tabular}{c|c|c|c|c|c|c|c|c|c}

\hline
\hline
\\
NIRSpec ID & RA & Dec & z & $\log{M_*/M_{\odot}}$ & $R_{\rm e}$ [kpc] & $\rm 12+log(O+H)_{\rm N2S2H\alpha}$ & $\rm 12+log(O+H)_{\rm N2}$ & $\rm 12+log(O+H)_{\rm R23}$ & $\rm 12+log(O+H)_{\rm O3N2}$ \\
\hline
110 & 189.146 & 62.216 & 4.064 & 9.268 $\pm$ 0.200 & 1.165 & 8.499 $\pm$ 0.054 & 8.263 $\pm$ 0.008 & 8.359 $\pm$ 0.027 & 8.184 $\pm$ 0.008 \\

804 & 189.134 & 62.242 & 3.662 & 9.356 $\pm$ 0.184 & 1.367 & 8.912 $\pm$ 0.093 & 8.318 $\pm$ 0.015 & 8.214 $\pm$ 0.086 & 8.260 $\pm$ 0.018 \\

917 & 189.080 & 62.255 & 4.400 & 9.075 $\pm$ 0.162 & 0.859 & - & 8.341 $\pm$ 0.023 & 8.521 $\pm$ 0.095 & 8.219 $\pm$ 0.022 \\

926 & 189.080 & 62.256 & 6.756 & 9.712 $\pm$ 0.206 & 1.434 & - & 8.383 $\pm$ 0.015 & 8.758 $\pm$ 0.034 & 8.312 $\pm$ 0.014 \\

1121 & 189.130 & 62.281 & 3.342 & 9.296 $\pm$ 0.182 & 1.082 & 8.642 $\pm$ 0.065 & 8.315 $\pm$ 0.009 & 8.417 $\pm$ 0.051 & 8.230 $\pm$ 0.010 \\

1233 & 189.117 & 62.298 & 3.333 & 9.728 $\pm$ 0.206 & 2.293 & 8.703 $\pm$ 0.084 & 8.379 $\pm$ 0.017 & 8.489 $\pm$ 0.104 & 8.298 $\pm$ 0.019 \\

1237 & 189.113 & 62.298 & 3.330 & 9.330 $\pm$ 0.202 & 1.706 & 8.698 $\pm$ 0.104 & 8.278 $\pm$ 0.020 & 8.252 $\pm$ 0.053 & 8.227 $\pm$ 0.018 \\

1927 & 189.071 & 62.278 & 3.660 & 9.349 $\pm$ 0.159 & 0.400 & 8.936 $\pm$ 0.094 & 8.314 $\pm$ 0.012 & 8.361 $\pm$ 0.055 & 8.222 $\pm$ 0.012 \\

2707 & 189.176 & 62.232 & 2.987 & 10.038 $\pm$ 0.193 & 2.268 & 8.826 $\pm$ 0.052 & 8.427 $\pm$ 0.010 & 8.977 $\pm$ 0.034 & 8.425 $\pm$ 0.017 \\

2945 & 189.141 & 62.281 & 3.335 & 9.667 $\pm$ 0.195 & 1.047 & 8.675 $\pm$ 0.049 & 8.335 $\pm$ 0.007 & 8.826 $\pm$ 0.018 & 8.291 $\pm$ 0.007 \\

3451 & 189.191 & 62.239 & 3.176 & 9.649 $\pm$ 0.146 & 2.261 & 8.530 $\pm$ 0.061 & 8.326 $\pm$ 0.010 & 8.307 $\pm$ 0.040 & 8.287 $\pm$ 0.010 \\

5670 & 189.180 & 62.242 & 2.207 & 8.814 $\pm$ 0.171 & 1.443 & 8.112 $\pm$ 0.131 & 8.190 $\pm$ 0.028 & - & 8.146 $\pm$ 0.024 \\

11836 & 189.221 & 62.264 & 4.409 & 8.471 $\pm$ 0.194 & 0.295 & - & 8.181 $\pm$ 0.014 & 7.996 $\pm$ 0.024 & 8.096 $\pm$ 0.011 \\

15529 & 189.215 & 62.277 & 3.871 & 9.066 $\pm$ 0.262 & 1.228 & 8.108 $\pm$ 0.127 & 8.109 $\pm$ 0.027 & 8.359 $\pm$ 0.033 & - \\

20185 & 189.107 & 62.231 & 2.931 & 9.531 $\pm$ 0.179 & 1.844 & 8.763 $\pm$ 0.078 & 8.300 $\pm$ 0.015 & 8.403 $\pm$ 0.054 & 8.274 $\pm$ 0.015 \\

21158 & 189.085 & 62.236 & 1.872 & 9.693 $\pm$ 0.133 & 3.522 & 8.333 $\pm$ 0.102 & 8.191 $\pm$ 0.024 & 8.209 $\pm$ 0.046 & 8.182 $\pm$ 0.021 \\

22411 & 189.041 & 62.243 & 3.681 & 8.957 $\pm$ 0.204 & 1.574 & - & 8.341 $\pm$ 0.032 & 8.507 $\pm$ 0.122 & 8.212 $\pm$ 0.030 \\

22481 & 189.107 & 62.243 & 3.335 & 9.697 $\pm$ 0.196 & 1.964 & 8.319 $\pm$ 0.100 & 8.263 $\pm$ 0.020 & 8.356 $\pm$ 0.055 & 8.228 $\pm$ 0.018 \\

22837 & 189.023 & 62.246 & 4.152 & 9.394 $\pm$ 0.166 & 0.811 & - & 8.368 $\pm$ 0.020 & 8.767 $\pm$ 0.047 & 8.269 $\pm$ 0.018 \\

23742 & 189.081 & 62.251 & 3.126 & 9.144 $\pm$ 0.232 & 1.317 & - & 8.227 $\pm$ 0.013 & 8.263 $\pm$ 0.032 & 8.157 $\pm$ 0.012 \\

24728 & 189.190 & 62.256 & 1.909 & 9.852 $\pm$ 0.106 & 1.778 & 8.444 $\pm$ 0.045 & 8.375 $\pm$ 0.009 & 8.541 $\pm$ 0.075 & 8.354 $\pm$ 0.014 \\

24755 & 189.148 & 62.256 & 2.981 & 8.902 $\pm$ 0.048 & 1.404 & - & 8.139 $\pm$ 0.031 & 8.096 $\pm$ 0.042 & 8.084 $\pm$ 0.025 \\

24835 & 189.124 & 62.225 & 2.933 & 9.664 $\pm$ 0.194 & 2.580 & 8.654 $\pm$ 0.099 & 8.346 $\pm$ 0.020 & 8.802 $\pm$ 0.059 & 8.337 $\pm$ 0.021 \\

25030 & 189.095 & 62.257 & 1.745 & 9.264 $\pm$ 0.121 & 1.238 & 8.465 $\pm$ 0.114 & 8.167 $\pm$ 0.027 & - & 8.097 $\pm$ 0.022 \\

25528 & 189.003 & 62.260 & 2.039 & 9.737 $\pm$ 0.083 & 3.684 & - & 8.381 $\pm$ 0.025 & 8.870 $\pm$ 0.062 & 8.393 $\pm$ 0.027 \\

25603 & 189.170 & 62.233 & 2.989 & 9.571 $\pm$ 0.188 & 1.165 & 8.264 $\pm$ 0.053 & 8.242 $\pm$ 0.011 & 8.308 $\pm$ 0.033 & 8.190 $\pm$ 0.010 \\

25766 & 189.178 & 62.234 & 1.989 & 9.774 $\pm$ 0.086 & 2.573 & 8.535 $\pm$ 0.068 & 8.397 $\pm$ 0.013 & 8.853 $\pm$ 0.054 & 8.416 $\pm$ 0.018 \\

25771 & 189.001 & 62.262 & 1.675 & 8.012 $\pm$ 0.157 & 0.690 & 8.089 $\pm$ 0.124 & 8.102 $\pm$ 0.026 & - & - \\

25844 & 189.114 & 62.235 & 2.273 & 9.317 $\pm$ 0.102 & 2.422 & 8.127 $\pm$ 0.126 & 8.241 $\pm$ 0.028 & 8.347 $\pm$ 0.068 & 8.218 $\pm$ 0.025 \\

26003 & 189.097 & 62.263 & 2.967 & 8.231 $\pm$ 0.103 & 0.774 & - & 8.278 $\pm$ 0.032 & - & 8.169 $\pm$ 0.028 \\

26113 & 189.100 & 62.264 & 3.226 & 8.945 $\pm$ 0.126 & 0.794 & 8.250 $\pm$ 0.094 & 8.182 $\pm$ 0.018 & 8.360 $\pm$ 0.033 & 8.112 $\pm$ 0.015 \\

26149 & 189.097 & 62.238 & 1.431 & 9.696 $\pm$ 0.100 & 3.872 & 8.439 $\pm$ 0.091 & 8.381 $\pm$ 0.017 & - & 8.414 $\pm$ 0.016 \\

26194 & 189.172 & 62.238 & 1.987 & 9.961 $\pm$ 0.069 & 1.948 & 8.760 $\pm$ 0.034 & 8.386 $\pm$ 0.004 & - & 8.371 $\pm$ 0.009 \\

26227 & 189.078 & 62.264 & 1.990 & 9.988 $\pm$ 0.105 & 1.641 & 8.674 $\pm$ 0.060 & 8.435 $\pm$ 0.011 & 9.065 $\pm$ 0.017 & 8.534 $\pm$ 0.022 \\

26276 & 189.224 & 62.238 & 2.975 & 9.818 $\pm$ 0.153 & 1.824 & 8.724 $\pm$ 0.066 & 8.396 $\pm$ 0.009 & 8.803 $\pm$ 0.044 & 8.360 $\pm$ 0.012 \\
\hline
\hline
\end{tabular}
\caption{The table containing all of galaxies used in this work. Extensive information related to the emission lines, after correcting for extinction, can be accessed through \href{https://drive.google.com/file/d/1_X__k1Adm_l-1Mz2Y-oq7b7gW-Gtk8TN/view?usp=share_link}{this link}.} \label{tab:data1}
\end{center}
\end{rotatetable*}
\end{table*}

\begin{table*}[h]
\movetabledown=2.0in
\begin{rotatetable*}
\begin{center}
\begin{tabular}{c|c|c|c|c|c|c|c|c|c}

\hline
\hline
\\
NIRSpec ID & RA & Dec & z & $\log{M_*/M_{\odot}}$ & $R_{\rm e}$ [kpc] & $\rm 12+log(O+H)_{\rm N2S2H\alpha}$ & $\rm 12+log(O+H)_{\rm N2}$ & $\rm 12+log(O+H)_{\rm R23}$ & $\rm 12+log(O+H)_{\rm O3N2}$ \\
\hline
26544 & 189.103 & 62.265 & 2.973 & 10.035 $\pm$ 0.216 & 2.272 & 8.590 $\pm$ 0.057 & 8.401 $\pm$ 0.010 & 8.754 $\pm$ 0.077 & 8.439 $\pm$ 0.018 \\

26751 & 189.008 & 62.267 & 2.176 & 8.482 $\pm$ 0.205 & 0.774 & 8.317 $\pm$ 0.081 & 8.174 $\pm$ 0.016 & 8.334 $\pm$ 0.029 & 8.097 $\pm$ 0.013 \\

27003 & 189.015 & 62.268 & 5.596 & 9.126 $\pm$ 0.230 & 1.018 & - & 8.304 $\pm$ 0.028 & 8.220 $\pm$ 0.083 & 8.195 $\pm$ 0.025 \\

27917 & 189.095 & 62.273 & 1.996 & 9.794 $\pm$ 0.128 & 0.904 & 8.830 $\pm$ 0.056 & 8.491 $\pm$ 0.009 & 8.988 $\pm$ 0.024 & 8.560 $\pm$ 0.015 \\

28139 & 189.189 & 62.251 & 1.530 & 9.451 $\pm$ 0.079 & 1.646 & - & 8.375 $\pm$ 0.033 & - & 8.405 $\pm$ 0.028 \\

28174 & 189.228 & 62.252 & 4.402 & 9.079 $\pm$ 0.188 & 1.530 & 8.223 $\pm$ 0.084 & 8.170 $\pm$ 0.017 & 8.334 $\pm$ 0.029 & 8.105 $\pm$ 0.014 \\

28187 & 189.121 & 62.275 & 3.207 & 9.637 $\pm$ 0.155 & 1.024 & 8.551 $\pm$ 0.102 & 8.286 $\pm$ 0.023 & 8.309 $\pm$ 0.060 & 8.234 $\pm$ 0.020 \\

28271 & 189.068 & 62.275 & 2.093 & 9.297 $\pm$ 0.158 & 1.116 & 8.189 $\pm$ 0.038 & 8.169 $\pm$ 0.009 & 8.242 $\pm$ 0.016 & 8.129 $\pm$ 0.007 \\

28390 & 189.071 & 62.276 & 1.677 & 9.664 $\pm$ 0.169 & 1.504 & 8.348 $\pm$ 0.102 & 8.307 $\pm$ 0.021 & - & 8.323 $\pm$ 0.023 \\

28433 & 189.135 & 62.277 & 1.522 & 9.888 $\pm$ 0.115 & 3.974 & 8.637 $\pm$ 0.111 & 8.331 $\pm$ 0.024 & - & 8.349 $\pm$ 0.023 \\

29023 & 189.091 & 62.280 & 2.093 & 10.017 $\pm$ 0.086 & 1.722 & 8.798 $\pm$ 0.064 & 8.442 $\pm$ 0.009 & 8.274 $\pm$ 0.195 & 8.422 $\pm$ 0.023 \\

29034 & 189.094 & 62.280 & 2.093 & 9.918 $\pm$ 0.109 & 1.995 & 8.666 $\pm$ 0.065 & 8.432 $\pm$ 0.010 & 8.880 $\pm$ 0.079 & 8.498 $\pm$ 0.028 \\

29438 & 189.086 & 62.283 & 1.675 & 9.473 $\pm$ 0.215 & 2.040 & 8.425 $\pm$ 0.060 & 8.359 $\pm$ 0.021 & - & 8.282 $\pm$ 0.020 \\

29648 & 189.209 & 62.264 & 2.960 & 9.729 $\pm$ 0.178 & 3.652 & 8.698 $\pm$ 0.027 & 8.340 $\pm$ 0.004 & 8.808 $\pm$ 0.017 & 8.310 $\pm$ 0.005 \\

29715 & 189.092 & 62.284 & 1.879 & 9.534 $\pm$ 0.080 & 0.614 & 8.804 $\pm$ 0.085 & 8.369 $\pm$ 0.012 & 8.826 $\pm$ 0.028 & 8.350 $\pm$ 0.011 \\

30206 & 189.071 & 62.287 & 1.587 & 9.904 $\pm$ 0.159 & 1.044 & 8.978 $\pm$ 0.081 & 8.460 $\pm$ 0.007 & - & 8.517 $\pm$ 0.007 \\

31514 & 189.134 & 62.276 & 1.487 & 9.093 $\pm$ 0.189 & 1.898 & 8.518 $\pm$ 0.032 & 8.297 $\pm$ 0.006 & - & 8.240 $\pm$ 0.005 \\

31896 & 189.124 & 62.279 & 2.271 & 10.071 $\pm$ 0.141 & 1.949 & 9.081 $\pm$ 0.049 & 8.490 $\pm$ 0.005 & 8.823 $\pm$ 0.064 & 8.525 $\pm$ 0.018 \\

32332 & 189.150 & 62.283 & 2.485 & 9.953 $\pm$ 0.065 & 2.289 & 8.835 $\pm$ 0.116 & 8.385 $\pm$ 0.024 & 8.986 $\pm$ 0.039 & 8.360 $\pm$ 0.031 \\

32665 & 189.177 & 62.303 & 2.037 & 9.632 $\pm$ 0.082 & 1.407 & - & 8.419 $\pm$ 0.015 & 9.068 $\pm$ 0.021 & 8.514 $\pm$ 0.027 \\

42939 & 189.210 & 62.164 & 3.467 & 8.380 $\pm$ 0.197 & 0.402 & 8.583 $\pm$ 0.115 & 8.165 $\pm$ 0.024 & 8.315 $\pm$ 0.033 & - \\

44099 & 189.257 & 62.167 & 5.859 & 9.291 $\pm$ 0.178 & 0.792 & 8.539 $\pm$ 0.100 & 8.279 $\pm$ 0.022 & 8.221 $\pm$ 0.054 & 8.203 $\pm$ 0.019 \\

55967 & 189.221 & 62.200 & 2.086 & 8.941 $\pm$ 0.235 & 1.270 & 8.044 $\pm$ 0.078 & 8.179 $\pm$ 0.018 & 8.365 $\pm$ 0.038 & 8.134 $\pm$ 0.015 \\

59350 & 189.207 & 62.210 & 2.016 & 9.849 $\pm$ 0.098 & 2.235 & 8.699 $\pm$ 0.055 & 8.381 $\pm$ 0.009 & 8.943 $\pm$ 0.021 & 8.403 $\pm$ 0.011 \\

64685 & 189.019 & 62.293 & 2.210 & 9.443 $\pm$ 0.120 & 0.664 & 8.475 $\pm$ 0.061 & 8.309 $\pm$ 0.010 & 8.337 $\pm$ 0.052 & 8.267 $\pm$ 0.012 \\

64818 & 189.027 & 62.290 & 1.359 & 9.824 $\pm$ 0.154 & 0.515 & 9.094 $\pm$ 0.036 & 8.486 $\pm$ 0.003 & - & 8.466 $\pm$ 0.004 \\

64984 & 189.038 & 62.280 & 1.337 & 9.302 $\pm$ 0.145 & 0.816 & - & 8.338 $\pm$ 0.015 & - & 8.334 $\pm$ 0.013 \\ 

65062 & 189.042 & 62.297 & 1.484 & 9.910 $\pm$ 0.229 & 3.992 & 8.545 $\pm$ 0.045 & 8.377 $\pm$ 0.009 & - & 8.440 $\pm$ 0.008 \\

65099 & 189.044 & 62.301 & 2.811 & 9.570 $\pm$ 0.186 & 1.894 & 8.726 $\pm$ 0.072 & 8.273 $\pm$ 0.015 & 8.707 $\pm$ 0.095 & 8.193 $\pm$ 0.017 \\

71093 & 189.276 & 62.162 & 5.035 & 9.890 $\pm$ 0.155 & 2.016 & - & 8.308 $\pm$ 0.013 & 8.016 $\pm$ 0.050 & 8.292 $\pm$ 0.014 \\

71890 & 189.276 & 62.167 & 3.414 & 9.032 $\pm$ 0.118 & 0.629 & - & 8.352 $\pm$ 0.017 & 8.565 $\pm$ 0.033 & 8.257 $\pm$ 0.014 \\

73357 & 189.222 & 62.177 & 1.978 & 9.542 $\pm$ 0.119 & 1.712 & 8.435 $\pm$ 0.032 & 8.282 $\pm$ 0.007 & - & 8.230 $\pm$ 0.006 \\

76983 & 189.261 & 62.195 & 2.851 & 8.853 $\pm$ 0.181 & 0.655 & 8.585 $\pm$ 0.080 & 8.215 $\pm$ 0.016 & 8.340 $\pm$ 0.060 & 8.111 $\pm$ 0.015 \\

78773 & 189.203 & 62.205 & 5.188 & 9.364 $\pm$ 0.040 & 1.815 & 8.398 $\pm$ 0.055 & 8.230 $\pm$ 0.010 & 8.199 $\pm$ 0.026 & 8.146 $\pm$ 0.009 \\

78888 & 189.275 & 62.205 & 3.676 & 9.317 $\pm$ 0.116 & 1.647 & 8.706 $\pm$ 0.111 & 8.263 $\pm$ 0.023 & 8.378 $\pm$ 0.051 & 8.177 $\pm$ 0.019 \\
\\
\hline
\hline
\end{tabular}
\caption{Continuation of Table \ref{tab:data1}} \label{tab:data2}
\end{center}
\end{rotatetable*}
\end{table*}

\begin{table*}[h]
\movetabledown=2.0in
\begin{rotatetable*}
\begin{center}
\begin{tabular}{c|c|c|c|c|c|c|c|c|c}

\hline
\hline
\\
NIRSpec ID & RA & Dec & z & $\log{M_*/M_{\odot}}$ & $R_{\rm e}$ [kpc] & $\rm 12+log(O+H)_{\rm N2S2H\alpha}$ & $\rm 12+log(O+H)_{\rm N2}$ & $\rm 12+log(O+H)_{\rm R23}$ & $\rm 12+log(O+H)_{\rm O3N2}$ \\
\hline
79203 & 189.307 & 62.206 & 2.270 & 9.619 $\pm$ 0.136 & 2.456 & 8.573 $\pm$ 0.077 & 8.361 $\pm$ 0.013 & 8.592 $\pm$ 0.082 & 8.349 $\pm$ 0.017 \\

79309 & 189.225 & 62.207 & 1.796 & 9.591 $\pm$ 0.090 & 1.920 & 8.352 $\pm$ 0.045 & 8.282 $\pm$ 0.009 & 8.252 $\pm$ 0.032 & 8.241 $\pm$ 0.009 \\

80742 & 189.218 & 62.214 & 2.420 & 8.940 $\pm$ 0.132 & 1.177 & 8.762 $\pm$ 0.100 & 8.253 $\pm$ 0.015 & 8.444 $\pm$ 0.068 & 8.164 $\pm$ 0.015 \\

81621 & 189.228 & 62.221 & 2.052 & 9.534 $\pm$ 0.106 & 0.879 & 8.260 $\pm$ 0.063 & 8.265 $\pm$ 0.014 & 8.416 $\pm$ 0.038 & 8.215 $\pm$ 0.013 \\

3753 & 53.112 & -27.817 & 1.768 & 9.376 $\pm$ 0.107 & 0.923 & 8.264 $\pm$ 0.039 & 8.243 $\pm$ 0.008 & 8.364 $\pm$ 0.018 & 8.196 $\pm$ 0.007 \\

3892 & 53.163 & -27.817 & 2.807 & 9.100 $\pm$ 0.068 & 1.243 & 8.138 $\pm$ 0.048 & 8.178 $\pm$ 0.010 & 8.288 $\pm$ 0.028 & 8.112 $\pm$ 0.009 \\

18970 & 53.156 & -27.772 & 3.725 & 8.953 $\pm$ 0.245 & 0.774 & 8.528 $\pm$ 0.104 & 8.250 $\pm$ 0.026 & 8.226 $\pm$ 0.045 & 8.183 $\pm$ 0.021 \\

34104 & 53.181 & -27.822 & 5.607 & 9.206 $\pm$ 0.220 & 0.405 & - & 8.226 $\pm$ 0.022 & 8.200 $\pm$ 0.043 & 8.137 $\pm$ 0.018 \\

36679 & 53.138 & -27.816 & 2.617 & 9.948 $\pm$ 0.170 & 2.621 & 8.577 $\pm$ 0.081 & 8.378 $\pm$ 0.015 & 8.981 $\pm$ 0.056 & 8.512 $\pm$ 0.034 \\

39898 & 53.132 & -27.809 & 3.170 & 9.864 $\pm$ 0.156 & 2.742 & 8.690 $\pm$ 0.100 & 8.303 $\pm$ 0.018 & 8.558 $\pm$ 0.076 & 8.266 $\pm$ 0.019 \\

45976 & 53.153 & -27.794 & 3.084 & 9.380 $\pm$ 0.200 & 1.066 & 8.253 $\pm$ 0.054 & 8.242 $\pm$ 0.011 & - & 8.184 $\pm$ 0.010 \\

47870 & 53.149 & -27.789 & 1.908 & 9.954 $\pm$ 0.188 & 2.411 & 8.582 $\pm$ 0.036 & 8.359 $\pm$ 0.006 & 8.787 $\pm$ 0.015 & 8.376 $\pm$ 0.006 \\

48489 & 53.195 & -27.787 & 3.645 & 9.164 $\pm$ 0.121 & 1.194 & - & 8.202 $\pm$ 0.031 & 8.211 $\pm$ 0.055 & 8.117 $\pm$ 0.026 \\

48912 & 53.192 & -27.786 & 2.454 & 9.438 $\pm$ 0.112 & 1.300 & 8.439 $\pm$ 0.090 & 8.238 $\pm$ 0.015 & 8.069 $\pm$ 0.067 & 8.209 $\pm$ 0.017 \\

49344 & 53.149 & -27.785 & 2.070 & 9.274 $\pm$ 0.108 & 1.986 & - & 8.309 $\pm$ 0.027 & 8.771 $\pm$ 0.154 & 8.236 $\pm$ 0.030 \\

54612 & 53.145 & -27.771 & 3.083 & 8.915 $\pm$ 0.113 & 0.889 & 8.314 $\pm$ 0.048 & 8.201 $\pm$ 0.008 & 8.258 $\pm$ 0.021 & 8.141 $\pm$ 0.007 \\

58112 & 53.141 & -27.762 & 1.622 & 10.044 $\pm$ 0.170 & 2.343 & 8.852 $\pm$ 0.059 & 8.461 $\pm$ 0.008 & - & 8.608 $\pm$ 0.020 \\

61321 & 53.154 & -27.752 & 4.842 & 8.336 $\pm$ 0.066 & 0.255 & - & 8.248 $\pm$ 0.024 & 8.316 $\pm$ 0.119 & 8.122 $\pm$ 0.025 \\

62328 & 53.149 & -27.749 & 2.573 & 9.247 $\pm$ 0.183 & 1.724 & 8.324 $\pm$ 0.044 & 8.232 $\pm$ 0.009 & 8.248 $\pm$ 0.030 & 8.156 $\pm$ 0.008 \\

120414 & 53.164 & -27.787 & 1.095 & 8.847 $\pm$ 0.204 & 0.804 & 8.140 $\pm$ 0.051 & 8.183 $\pm$ 0.011 & - & 8.116 $\pm$ 0.009 \\

210262 & 53.196 & -27.773 & 2.820 & 9.315 $\pm$ 0.076 & 0.992 & - & 8.234 $\pm$ 0.027 & 8.346 $\pm$ 0.088 & 8.161 $\pm$ 0.025 \\

210343 & 53.197 & -27.772 & 2.811 & 8.793 $\pm$ 0.061 & 1.636 & - & 8.219 $\pm$ 0.030 & 8.376 $\pm$ 0.100 & 8.115 $\pm$ 0.027 \\

227330 & 53.196 & -27.825 & 2.084 & 9.145 $\pm$ 0.136 & 0.894 & 8.317 $\pm$ 0.064 & 8.239 $\pm$ 0.012 & 8.320 $\pm$ 0.037 & 8.173 $\pm$ 0.011 \\

10007444 & 53.149 & -27.782 & 1.906 & 9.992 $\pm$ 0.105 & 4.186 & 8.278 $\pm$ 0.067 & 8.312 $\pm$ 0.016 & 8.375 $\pm$ 0.060 & 8.318 $\pm$ 0.016 \\

10008071 & 53.154 & -27.771 & 2.227 & 9.578 $\pm$ 0.158 & 3.247 & 8.257 $\pm$ 0.031 & 8.271 $\pm$ 0.007 & 8.411 $\pm$ 0.026 & 8.231 $\pm$ 0.007 \\

10013704 & 53.127 & -27.818 & 5.920 & 8.790 $\pm$ 0.125 & 0.895 & - & 8.346 $\pm$ 0.019 & - & 8.245 $\pm$ 0.017 \\

10033232 & 53.170 & -27.834 & 1.781 & 9.474 $\pm$ 0.073 & 2.830 & 8.164 $\pm$ 0.087 & 8.205 $\pm$ 0.019 & 8.371 $\pm$ 0.031 & 8.169 $\pm$ 0.016 \\
\hline
\hline
\end{tabular}
\caption{Continuation of Table \ref{tab:data1} and Table \ref{tab:data2}} \label{tab:data3}
\end{center}
\end{rotatetable*}
\end{table*}

\bibliography{ref}

\begin{thebibliography}{}
\expandafter\ifx\csname natexlab\endcsname\relax\def\natexlab#1{#1}\fi
\providecommand{\url}[1]{\href{#1}{#1}}
\providecommand{\dodoi}[1]{doi:~\href{http://doi.org/#1}{\nolinkurl{#1}}}
\providecommand{\doeprint}[1]{\href{http://ascl.net/#1}{\nolinkurl{http://ascl.net/#1}}}
\providecommand{\doarXiv}[1]{\href{https://arxiv.org/abs/#1}{\nolinkurl{https://arxiv.org/abs/#1}}}

\bibitem[{{Bian} {et~al.}(2018){Bian}, {Kewley}, \& {Dopita}}]{Bian_2018}
{Bian}, F., {Kewley}, L.~J., \& {Dopita}, M.~A. 2018, \apj, 859, 175, \dodoi{10.3847/1538-4357/aabd74}

\bibitem[{{Boquien}(2020)}]{Boquien_2020}
{Boquien}, M. 2020, in American Astronomical Society Meeting Abstracts, Vol. 235, American Astronomical Society Meeting Abstracts \#235, 228.01

\bibitem[{{Bouch{\'e}} {et~al.}(2010){Bouch{\'e}}, {Dekel}, {Genzel}, {Genel}, {Cresci}, {F{\"o}rster Schreiber}, {Shapiro}, {Davies}, \& {Tacconi}}]{Bouche_2010}
{Bouch{\'e}}, N., {Dekel}, A., {Genzel}, R., {et~al.} 2010, \apj, 718, 1001, \dodoi{10.1088/0004-637X/718/2/1001}

\bibitem[{Brammer(2023)}]{_zenodo}
Brammer, G. 2023, grizli,  Zenodo, \dodoi{10.5281/ZENODO.1146904}

\bibitem[{{Bruzual} \& {Charlot}(2003)}]{Bruzual_Charlot_2003}
{Bruzual}, G., \& {Charlot}, S. 2003, \mnras, 344, 1000, \dodoi{10.1046/j.1365-8711.2003.06897.x}

\bibitem[{{Bundy} {et~al.}(2015){Bundy}, {Bershady}, {Law}, {Yan}, {Drory}, {MacDonald}, {Wake}, {Cherinka}, {S{\'a}nchez-Gallego}, {Weijmans}, {Thomas}, {Tremonti}, {Masters}, {Coccato}, {Diamond-Stanic}, {Arag{\'o}n-Salamanca}, {Avila-Reese}, {Badenes}, {Falc{\'o}n-Barroso}, {Belfiore}, {Bizyaev}, {Blanc}, {Bland-Hawthorn}, {Blanton}, {Brownstein}, {Byler}, {Cappellari}, {Conroy}, {Dutton}, {Emsellem}, {Etherington}, {Frinchaboy}, {Fu}, {Gunn}, {Harding}, {Johnston}, {Kauffmann}, {Kinemuchi}, {Klaene}, {Knapen}, {Leauthaud}, {Li}, {Lin}, {Maiolino}, {Malanushenko}, {Malanushenko}, {Mao}, {Maraston}, {McDermid}, {Merrifield}, {Nichol}, {Oravetz}, {Pan}, {Parejko}, {Sanchez}, {Schlegel}, {Simmons}, {Steele}, {Steinmetz}, {Thanjavur}, {Thompson}, {Tinker}, {van den Bosch}, {Westfall}, {Wilkinson}, {Wright}, {Xiao}, \& {Zhang}}]{Bundy-15}
{Bundy}, K., {Bershady}, M.~A., {Law}, D.~R., {et~al.} 2015, \apj, 798, 7, \dodoi{10.1088/0004-637X/798/1/7}

\bibitem[{{Calzetti} {et~al.}(2000){Calzetti}, {Armus}, {Bohlin}, {Kinney}, {Koornneef}, \& {Storchi-Bergmann}}]{Calzetti_2000}
{Calzetti}, D., {Armus}, L., {Bohlin}, R.~C., {et~al.} 2000, \apj, 533, 682, \dodoi{10.1086/308692}

\bibitem[{Carnall {et~al.}(2018)Carnall, McLure, Dunlop, \& Davé}]{Carnall_2018}
Carnall, A.~C., McLure, R.~J., Dunlop, J.~S., \& Davé, R. 2018, Monthly Notices of the Royal Astronomical Society, 480, 4379–4401, \dodoi{10.1093/mnras/sty2169}

\bibitem[{Chakraborty {et~al.}(2024)Chakraborty, Sarkar, Smith, Ferland, McDonald, Forman, Vogelsberger, Torrey, Garcia, Bautz, Foster, Miller, \& Grant}]{Chakraborty_2024}
Chakraborty, P., Sarkar, A., Smith, R., {et~al.} 2024, Unveiling the Cosmic Chemistry II: "direct" $T_e$-based metallicity of galaxies at 3 $< z <$ 10 with JWST/NIRSpec.
\newblock \doarXiv{2412.15435}

\bibitem[{Chemerynska {et~al.}(2024)Chemerynska, Atek, Dayal, Furtak, Feldmann, Greene, Maseda, Nanayakkara, Oesch, Fujimoto, Labbé, Bezanson, Brammer, Cutler, Leja, Pan, Price, Wang, Weaver, \& Whitaker}]{Chemerynska_2024}
Chemerynska, I., Atek, H., Dayal, P., {et~al.} 2024, The Astrophysical Journal Letters, 976, L15, \dodoi{10.3847/2041-8213/ad8dc9}

\bibitem[{{Chen} {et~al.}(2025){Chen}, {Wang}, {Zou}, {Zou}, {Gao}, {Wang}, {Yu}, {Jia}, {Li}, {Ma}, {Yao}, {Ding}, \& {Zhu}}]{Chen25}
{Chen}, Z., {Wang}, E., {Zou}, H., {et~al.} 2025, \apj, 981, 81, \dodoi{10.3847/1538-4357/ada942}

\bibitem[{Christensen {et~al.}(2018)Christensen, Davé, Brooks, Quinn, \& Shen}]{Christensen_2018}
Christensen, C.~R., Davé, R., Brooks, A., Quinn, T., \& Shen, S. 2018, The Astrophysical Journal, 867, 142, \dodoi{10.3847/1538-4357/aae374}

\bibitem[{Cuestas {et~al.}(2025)Cuestas, Strom, Miller, Steidel, Trainor, Rudie, \& Nuñez}]{Cuestas_2025}
Cuestas, N. A.~K., Strom, A.~L., Miller, T.~B., {et~al.} 2025, Exploring the Relationship Between Stellar Mass, Metallicity, and Star Formation Rate at $z \sim 2.3$ in KBSS-MOSFIRE.
\newblock \doarXiv{2503.10800}

\bibitem[{{Curti} {et~al.}(2020){Curti}, {Mannucci}, {Cresci}, \& {Maiolino}}]{Curti_2020}
{Curti}, M., {Mannucci}, F., {Cresci}, G., \& {Maiolino}, R. 2020, \mnras, 491, 944, \dodoi{10.1093/mnras/stz2910}

\bibitem[{{Curti} {et~al.}(2024){Curti}, {Maiolino}, {Curtis-Lake}, {Chevallard}, {Carniani}, {D'Eugenio}, {Looser}, {Scholtz}, {Charlot}, {Cameron}, {{\"U}bler}, {Witstok}, {Boyett}, {Laseter}, {Sandles}, {Arribas}, {Bunker}, {Giardino}, {Maseda}, {Rawle}, {Rodr{\'\i}guez Del Pino}, {Smit}, {Willott}, {Eisenstein}, {Hausen}, {Johnson}, {Rieke}, {Robertson}, {Tacchella}, {Williams}, {Willmer}, {Baker}, {Bhatawdekar}, {Egami}, {Helton}, {Ji}, {Kumari}, {Perna}, {Shivaei}, \& {Sun}}]{Curti_2024}
{Curti}, M., {Maiolino}, R., {Curtis-Lake}, E., {et~al.} 2024, \aap, 684, A75, \dodoi{10.1051/0004-6361/202346698}

\bibitem[{{D'Eugenio} {et~al.}(2018){D'Eugenio}, {Colless}, {Groves}, {Bian}, \& {Barone}}]{DEugenio_2018}
{D'Eugenio}, F., {Colless}, M., {Groves}, B., {Bian}, F., \& {Barone}, T.~M. 2018, \mnras, 479, 1807, \dodoi{10.1093/mnras/sty1424}

\bibitem[{D'Eugenio {et~al.}(2024)D'Eugenio, Cameron, Scholtz, Carniani, Willott, Curtis-Lake, Bunker, Parlanti, Maiolino, Willmer, Jakobsen, Robertson, Johnson, Tacchella, Cargile, Rawle, Arribas, Chevallard, Curti, Egami, Eisenstein, Kumari, Looser, Rieke, Pino, Saxena, Übler, Venturi, Witstok, Baker, Bhatawdekar, Bonaventura, Boyett, Charlot, Danhaive, Hainline, Hausen, Helton, Ji, Ji, Jones, Joudžbalis, Maseda, Pérez-González, Perna, Puskás, Shivaei, Silcock, Simmonds, Smit, Sun, Villanueva, Williams, \& Zhu}]{DEugenio_2024}
D'Eugenio, F., Cameron, A.~J., Scholtz, J., {et~al.} 2024, JADES Data Release 3 -- NIRSpec/MSA spectroscopy for 4,000 galaxies in the GOODS fields.
\newblock \doarXiv{2404.06531}

\bibitem[{Dopita {et~al.}(2016)Dopita, Kewley, Sutherland, \& Nicholls}]{Dopita_2016}
Dopita, M.~A., Kewley, L.~J., Sutherland, R.~S., \& Nicholls, D.~C. 2016, Astrophysics and Space Science, 361, \dodoi{10.1007/s10509-016-2657-8}

\bibitem[{Easeman {et~al.}(2023)Easeman, Schady, Wuyts, \& Yates}]{Easeman_2023}
Easeman, B., Schady, P., Wuyts, S., \& Yates, R. 2023, Optimal metallicity diagnostics for MUSE observations of low-z galaxies.
\newblock \doarXiv{2311.03514}

\bibitem[{Eisenstein {et~al.}(2023{\natexlab{a}})Eisenstein, Johnson, Robertson, Tacchella, Hainline, Jakobsen, Maiolino, Bonaventura, Bunker, Cameron, Cargile, Curtis-Lake, Hausen, Puskás, Rieke, Sun, Willmer, Willott, Alberts, Arribas, Baker, Baum, Bhatawdekar, Carniani, Charlot, Chen, Chevallard, Curti, DeCoursey, D'Eugenio, de~Graaff, Egami, Helton, Ji, Jones, Kumari, Lützgendorf, Laseter, Looser, Lyu, Maseda, Nelson, Parlanti, Rauscher, Rawle, Rieke, Rix, Rujopakarn, Sandles, Saxena, Scholtz, Sharpe, Shivaei, Simmonds, Smit, Topping, Übler, Venturi, Williams, Witstok, \& Woodrum}]{Eisenstein_2023a}
Eisenstein, D.~J., Johnson, B.~D., Robertson, B., {et~al.} 2023{\natexlab{a}}, The JADES Origins Field: A New JWST Deep Field in the JADES Second NIRCam Data Release.
\newblock \doarXiv{2310.12340}

\bibitem[{Eisenstein {et~al.}(2023{\natexlab{b}})Eisenstein, Willott, Alberts, Arribas, Bonaventura, Bunker, Cameron, Carniani, Charlot, Curtis-Lake, D'Eugenio, Endsley, Ferruit, Giardino, Hainline, Hausen, Jakobsen, Johnson, Maiolino, Rieke, Rieke, Rix, Robertson, Stark, Tacchella, Williams, Willmer, Baker, Baum, Bhatawdekar, Boyett, Chen, Chevallard, Circosta, Curti, Danhaive, DeCoursey, de~Graaff, Dressler, Egami, Helton, Hviding, Ji, Jones, Kumari, Lützgendorf, Laseter, Looser, Lyu, Maseda, Nelson, Parlanti, Perna, Puskás, Rawle, Pino, Sandles, Saxena, Scholtz, Sharpe, Shivaei, Silcock, Simmonds, Skarbinski, Smit, Stone, Suess, Sun, Tang, Topping, Übler, Villanueva, Wallace, Whitler, Witstok, \& Woodrum}]{Eisenstein_2023b}
Eisenstein, D.~J., Willott, C., Alberts, S., {et~al.} 2023{\natexlab{b}}, Overview of the JWST Advanced Deep Extragalactic Survey (JADES).
\newblock \doarXiv{2306.02465}

\bibitem[{{Ellison} {et~al.}(2008){Ellison}, {Patton}, {Simard}, \& {McConnachie}}]{Ellison_2008}
{Ellison}, S.~L., {Patton}, D.~R., {Simard}, L., \& {McConnachie}, A.~W. 2008, \apjl, 672, L107, \dodoi{10.1086/527296}

\bibitem[{{Huang} {et~al.}(2019){Huang}, {Zou}, {Kong}, {Comparat}, {Lin}, {Gao}, {Liang}, {Delubac}, {Raichoor}, {Kneib}, {Schneider}, {Zhou}, {Yuan}, \& {Bershady}}]{Huang_2019}
{Huang}, C., {Zou}, H., {Kong}, X., {et~al.} 2019, \apj, 886, 31, \dodoi{10.3847/1538-4357/ab4902}

\bibitem[{{Inoue}(2011)}]{Inoue_2011}
{Inoue}, A.~K. 2011, \mnras, 415, 2920, \dodoi{10.1111/j.1365-2966.2011.18906.x}

\bibitem[{Jia {et~al.}(2024)Jia, Wang, Wang, Li, Yao, Song, Zhang, Rong, Chen, Yu, Chen, Li, Ma, \& Kong}]{Jia_2024}
Jia, C., Wang, E., Wang, H., {et~al.} 2024, Size Growth on Short Timescales of Star-Forming Galaxies: Insights from Size Variation with Rest-Frame Wavelength with JADES.
\newblock \doarXiv{2411.07458}

\bibitem[{{Juneau} {et~al.}(2014){Juneau}, {Bournaud}, {Charlot}, {Daddi}, {Elbaz}, {Trump}, {Brinchmann}, {Dickinson}, {Duc}, {Gobat}, {Jean-Baptiste}, {Le Floc'h}, {Lehnert}, {Pacifici}, {Pannella}, \& {Schreiber}}]{Juneau_2014}
{Juneau}, S., {Bournaud}, F., {Charlot}, S., {et~al.} 2014, \apj, 788, 88, \dodoi{10.1088/0004-637X/788/1/88}

\bibitem[{{Kewley} \& {Dopita}(2002)}]{Kewley_2002}
{Kewley}, L.~J., \& {Dopita}, M.~A. 2002, \apjs, 142, 35, \dodoi{10.1086/341326}

\bibitem[{{Kewley} \& {Ellison}(2008)}]{Kewley_2008}
{Kewley}, L.~J., \& {Ellison}, S.~L. 2008, \apj, 681, 1183, \dodoi{10.1086/587500}

\bibitem[{{Kobulnicky} \& {Kewley}(2004)}]{Kobulnicky_2004}
{Kobulnicky}, H.~A., \& {Kewley}, L.~J. 2004, \apj, 617, 240, \dodoi{10.1086/425299}

\bibitem[{{Langeroodi} \& {Hjorth}(2023)}]{Langeroodi_2023_1}
{Langeroodi}, D., \& {Hjorth}, J. 2023, arXiv e-prints, arXiv:2307.06336, \dodoi{10.48550/arXiv.2307.06336}

\bibitem[{Langeroodi {et~al.}(2023)Langeroodi, Hjorth, Chen, Kelly, Williams, Lin, Scarlata, Zitrin, Broadhurst, Diego, Huang, Filippenko, Foley, Jha, Koekemoer, Oguri, Perez-Fournon, Pierel, Poidevin, \& Strolger}]{Langeroodi_2023}
Langeroodi, D., Hjorth, J., Chen, W., {et~al.} 2023, The Astrophysical Journal, 957, 39, \dodoi{10.3847/1538-4357/acdbc1}

\bibitem[{{Lara-L{\'o}pez} {et~al.}(2010){Lara-L{\'o}pez}, {Cepa}, {Bongiovanni}, {P{\'e}rez Garc{\'\i}a}, {Ederoclite}, {Casta{\~n}eda}, {Fern{\'a}ndez Lorenzo}, {Povi{\'c}}, \& {S{\'a}nchez-Portal}}]{Lara_2010}
{Lara-L{\'o}pez}, M.~A., {Cepa}, J., {Bongiovanni}, A., {et~al.} 2010, \aap, 521, L53, \dodoi{10.1051/0004-6361/201014803}

\bibitem[{{Lequeux} {et~al.}(1979){Lequeux}, {Peimbert}, {Rayo}, {Serrano}, \& {Torres-Peimbert}}]{Lequeux_1979}
{Lequeux}, J., {Peimbert}, M., {Rayo}, J.~F., {Serrano}, A., \& {Torres-Peimbert}, S. 1979, \aap, 80, 155

\bibitem[{{Li} {et~al.}(2025){Li}, {Wang}, {Lyu}, {Chen}, {Wang}, {Chen}, {Yu}, {Jia}, \& {Ma}}]{Li-25}
{Li}, H., {Wang}, E., {Lyu}, C., {et~al.} 2025, arXiv e-prints, arXiv:2502.18785, \dodoi{10.48550/arXiv.2502.18785}

\bibitem[{{Lilly} {et~al.}(2013){Lilly}, {Carollo}, {Pipino}, {Renzini}, \& {Peng}}]{Lilly_2013}
{Lilly}, S.~J., {Carollo}, C.~M., {Pipino}, A., {Renzini}, A., \& {Peng}, Y. 2013, \apj, 772, 119, \dodoi{10.1088/0004-637X/772/2/119}

\bibitem[{Luo {et~al.}(2016)Luo, Brandt, Xue, Lehmer, Alexander, Bauer, Vito, Yang, Basu-Zych, Comastri, Gilli, Gu, Hornschemeier, Koekemoer, Liu, Mainieri, Paolillo, Ranalli, Rosati, Schneider, Shemmer, Smail, Sun, Tozzi, Vignali, \& Wang}]{Luo_2016}
Luo, B., Brandt, W.~N., Xue, Y.~Q., {et~al.} 2016, The Astrophysical Journal Supplement Series, 228, 2, \dodoi{10.3847/1538-4365/228/1/2}

\bibitem[{{Lyu} {et~al.}(2025){Lyu}, {Wang}, {Zhang}, {Peng}, {Wang}, {Li}, {Ma}, {Yu}, {Chen}, {Jia}, \& {Kong}}]{Lyu-25}
{Lyu}, C., {Wang}, E., {Zhang}, H., {et~al.} 2025, \apjl, 981, L6, \dodoi{10.3847/2041-8213/adb4ed}

\bibitem[{Ma {et~al.}(2024)Ma, Wang, Wang, Peng, Jiang, Yu, Jia, Chen, Li, \& Kong}]{Ma_2024}
Ma, C., Wang, K., Wang, E., {et~al.} 2024, The Astrophysical Journal Letters, 971, L14, \dodoi{10.3847/2041-8213/ad675f}

\bibitem[{{Maiolino} {et~al.}(2008){Maiolino}, {Nagao}, {Grazian}, {Cocchia}, {Marconi}, {Mannucci}, {Cimatti}, {Pipino}, {Ballero}, {Calura}, {Chiappini}, {Fontana}, {Granato}, {Matteucci}, {Pastorini}, {Pentericci}, {Risaliti}, {Salvati}, \& {Silva}}]{Maiolino_2008}
{Maiolino}, R., {Nagao}, T., {Grazian}, A., {et~al.} 2008, \aap, 488, 463, \dodoi{10.1051/0004-6361:200809678}

\bibitem[{{Mannucci} {et~al.}(2010){Mannucci}, {Cresci}, {Maiolino}, {Marconi}, \& {Gnerucci}}]{Mannucci_2010}
{Mannucci}, F., {Cresci}, G., {Maiolino}, R., {Marconi}, A., \& {Gnerucci}, A. 2010, \mnras, 408, 2115, \dodoi{10.1111/j.1365-2966.2010.17291.x}

\bibitem[{{Mannucci} {et~al.}(2009){Mannucci}, {Cresci}, {Maiolino}, {Marconi}, {Pastorini}, {Pozzetti}, {Gnerucci}, {Risaliti}, {Schneider}, {Lehnert}, \& {Salvati}}]{Mannucci_2009}
{Mannucci}, F., {Cresci}, G., {Maiolino}, R., {et~al.} 2009, \mnras, 398, 1915, \dodoi{10.1111/j.1365-2966.2009.15185.x}

\bibitem[{{Mitchell} {et~al.}(2018){Mitchell}, {Blaizot}, {Devriendt}, {Kimm}, {Michel-Dansac}, {Rosdahl}, \& {Slyz}}]{Mitchell_2018}
{Mitchell}, P.~D., {Blaizot}, J., {Devriendt}, J., {et~al.} 2018, \mnras, 474, 4279, \dodoi{10.1093/mnras/stx3017}

\bibitem[{{Muratov} {et~al.}(2017){Muratov}, {Kere{\v{s}}}, {Faucher-Gigu{\`e}re}, {Hopkins}, {Ma}, {Angl{\'e}s-Alc{\'a}zar}, {Chan}, {Torrey}, {Hafen}, {Quataert}, \& {Murray}}]{Muratov_2017}
{Muratov}, A.~L., {Kere{\v{s}}}, D., {Faucher-Gigu{\`e}re}, C.-A., {et~al.} 2017, \mnras, 468, 4170, \dodoi{10.1093/mnras/stx667}

\bibitem[{Nakajima {et~al.}(2023)Nakajima, Ouchi, Isobe, Harikane, Zhang, Ono, Umeda, \& Oguri}]{Nakajima_2023}
Nakajima, K., Ouchi, M., Isobe, Y., {et~al.} 2023, JWST Census for the Mass-Metallicity Star-Formation Relations at z=4-10 with the Self-Consistent Flux Calibration and the Proper Metallicity Calibrators.
\newblock \doarXiv{2301.12825}

\bibitem[{{Nakajima} {et~al.}(2012){Nakajima}, {Ouchi}, {Shimasaku}, {Ono}, {Lee}, {Foucaud}, {Ly}, {Dale}, {Salim}, {Finn}, {Almaini}, \& {Okamura}}]{Nakajima_2012}
{Nakajima}, K., {Ouchi}, M., {Shimasaku}, K., {et~al.} 2012, \apj, 745, 12, \dodoi{10.1088/0004-637X/745/1/12}

\bibitem[{{Nelson} {et~al.}(2018){Nelson}, {Pillepich}, {Springel}, {Weinberger}, {Hernquist}, {Pakmor}, {Genel}, {Torrey}, {Vogelsberger}, {Kauffmann}, {Marinacci}, \& {Naiman}}]{Nelson-18}
{Nelson}, D., {Pillepich}, A., {Springel}, V., {et~al.} 2018, \mnras, 475, 624, \dodoi{10.1093/mnras/stx3040}

\bibitem[{{O'Donnell}(1994)}]{ODonnell_1994}
{O'Donnell}, J.~E. 1994, \apj, 422, 158, \dodoi{10.1086/173713}

\bibitem[{{Peng} {et~al.}(2002){Peng}, {Ho}, {Impey}, \& {Rix}}]{Peng_2002}
{Peng}, C.~Y., {Ho}, L.~C., {Impey}, C.~D., \& {Rix}, H.-W. 2002, \aj, 124, 266, \dodoi{10.1086/340952}

\bibitem[{{Peng} {et~al.}(2010){Peng}, {Ho}, {Impey}, \& {Rix}}]{Peng_2010}
---. 2010, \aj, 139, 2097, \dodoi{10.1088/0004-6256/139/6/2097}

\bibitem[{{Peng} \& {Maiolino}(2014)}]{Peng_2014}
{Peng}, Y.-j., \& {Maiolino}, R. 2014, \mnras, 443, 3643, \dodoi{10.1093/mnras/stu1288}

\bibitem[{{Pillepich} {et~al.}(2018){Pillepich}, {Springel}, {Nelson}, {Genel}, {Naiman}, {Pakmor}, {Hernquist}, {Torrey}, {Vogelsberger}, {Weinberger}, \& {Marinacci}}]{Pillepich-18}
{Pillepich}, A., {Springel}, V., {Nelson}, D., {et~al.} 2018, \mnras, 473, 4077, \dodoi{10.1093/mnras/stx2656}

\bibitem[{{Planck Collaboration} {et~al.}(2016){Planck Collaboration}, {Ade}, {Aghanim}, {Arnaud}, {Ashdown}, {Aumont}, {Baccigalupi}, {Banday}, {Barreiro}, {Bartlett}, {Bartolo}, {Battaner}, {Battye}, {Benabed}, {Beno{\^\i}t}, {Benoit-L{\'e}vy}, {Bernard}, {Bersanelli}, {Bielewicz}, {Bock}, {Bonaldi}, {Bonavera}, {Bond}, {Borrill}, {Bouchet}, {Boulanger}, {Bucher}, {Burigana}, {Butler}, {Calabrese}, {Cardoso}, {Catalano}, {Challinor}, {Chamballu}, {Chary}, {Chiang}, {Chluba}, {Christensen}, {Church}, {Clements}, {Colombi}, {Colombo}, {Combet}, {Coulais}, {Crill}, {Curto}, {Cuttaia}, {Danese}, {Davies}, {Davis}, {de Bernardis}, {de Rosa}, {de Zotti}, {Delabrouille}, {D{\'e}sert}, {Di Valentino}, {Dickinson}, {Diego}, {Dolag}, {Dole}, {Donzelli}, {Dor{\'e}}, {Douspis}, {Ducout}, {Dunkley}, {Dupac}, {Efstathiou}, {Elsner}, {En{\ss}lin}, {Eriksen}, {Farhang}, {Fergusson}, {Finelli}, {Forni}, {Frailis}, {Fraisse}, {Franceschi}, {Frejsel}, {Galeotta}, {Galli}, {Ganga}, {Gauthier}, {Gerbino}, {Ghosh}, {Giard},
  {Giraud-H{\'e}raud}, {Giusarma}, {Gjerl{\o}w}, {Gonz{\'a}lez-Nuevo}, {G{\'o}rski}, {Gratton}, {Gregorio}, {Gruppuso}, {Gudmundsson}, {Hamann}, {Hansen}, {Hanson}, {Harrison}, {Helou}, {Henrot-Versill{\'e}}, {Hern{\'a}ndez-Monteagudo}, {Herranz}, {Hildebrandt}, {Hivon}, {Hobson}, {Holmes}, {Hornstrup}, {Hovest}, {Huang}, {Huffenberger}, {Hurier}, {Jaffe}, {Jaffe}, {Jones}, {Juvela}, {Keih{\"a}nen}, {Keskitalo}, {Kisner}, {Kneissl}, {Knoche}, {Knox}, {Kunz}, {Kurki-Suonio}, {Lagache}, {L{\"a}hteenm{\"a}ki}, {Lamarre}, {Lasenby}, {Lattanzi}, {Lawrence}, {Leahy}, {Leonardi}, {Lesgourgues}, {Levrier}, {Lewis}, {Liguori}, {Lilje}, {Linden-V{\o}rnle}, {L{\'o}pez-Caniego}, {Lubin}, {Mac{\'\i}as-P{\'e}rez}, {Maggio}, {Maino}, {Mandolesi}, {Mangilli}, {Marchini}, {Maris}, {Martin}, {Martinelli}, {Mart{\'\i}nez-Gonz{\'a}lez}, {Masi}, {Matarrese}, {McGehee}, {Meinhold}, {Melchiorri}, {Melin}, {Mendes}, {Mennella}, {Migliaccio}, {Millea}, {Mitra}, {Miville-Desch{\^e}nes}, {Moneti}, {Montier}, {Morgante}, {Mortlock},
  {Moss}, {Munshi}, {Murphy}, {Naselsky}, {Nati}, {Natoli}, {Netterfield}, {N{\o}rgaard-Nielsen}, {Noviello}, {Novikov}, {Novikov}, {Oxborrow}, {Paci}, {Pagano}, {Pajot}, {Paladini}, {Paoletti}, {Partridge}, {Pasian}, {Patanchon}, {Pearson}, {Perdereau}, {Perotto}, {Perrotta}, {Pettorino}, {Piacentini}, {Piat}, {Pierpaoli}, {Pietrobon}, {Plaszczynski}, {Pointecouteau}, {Polenta}, {Popa}, {Pratt}, \& {Pr{\'e}zeau}}]{Planck_2016}
{Planck Collaboration}, {Ade}, P.~A.~R., {Aghanim}, N., {et~al.} 2016, \aap, 594, A13, \dodoi{10.1051/0004-6361/201525830}

\bibitem[{{Rieke} {et~al.}(2023){Rieke}, {Robertson}, {Tacchella}, {Hainline}, {Johnson}, {Hausen}, {Ji}, {Willmer}, {Eisenstein}, {Pusk{\'a}s}, {Alberts}, {Arribas}, {Baker}, {Baum}, {Bhatawdekar}, {Bonaventura}, {Boyett}, {Bunker}, {Cameron}, {Carniani}, {Charlot}, {Chevallard}, {Chen}, {Curti}, {Curtis-Lake}, {Danhaive}, {DeCoursey}, {Dressler}, {Egami}, {Endsley}, {Helton}, {Hviding}, {Kumari}, {Looser}, {Lyu}, {Maiolino}, {Maseda}, {Nelson}, {Rieke}, {Rix}, {Sandles}, {Saxena}, {Sharpe}, {Shivaei}, {Skarbinski}, {Smit}, {Stark}, {Stone}, {Suess}, {Sun}, {Topping}, {{\"U}bler}, {Villanueva}, {Wallace}, {Williams}, {Willott}, {Whitler}, {Witstok}, \& {Woodrum}}]{Rieke_2023}
{Rieke}, M.~J., {Robertson}, B., {Tacchella}, S., {et~al.} 2023, \apjs, 269, 16, \dodoi{10.3847/1538-4365/acf44d}

\bibitem[{{S{\'a}nchez Almeida} \& {Dalla Vecchia}(2018)}]{Sanchez_2018}
{S{\'a}nchez Almeida}, J., \& {Dalla Vecchia}, C. 2018, \apj, 859, 109, \dodoi{10.3847/1538-4357/aac086}

\bibitem[{{S{\'a}nchez-Menguiano} {et~al.}(2024{\natexlab{a}}){S{\'a}nchez-Menguiano}, {S{\'a}nchez}, {S{\'a}nchez Almeida}, \& {Mu{\~n}oz-Tu{\~n}{\'o}n}}]{Sanchez_2024a}
{S{\'a}nchez-Menguiano}, L., {S{\'a}nchez}, S.~F., {S{\'a}nchez Almeida}, J., \& {Mu{\~n}oz-Tu{\~n}{\'o}n}, C. 2024{\natexlab{a}}, \aap, 682, L11, \dodoi{10.1051/0004-6361/202348423}

\bibitem[{{S{\'a}nchez-Menguiano} {et~al.}(2024{\natexlab{b}}){S{\'a}nchez-Menguiano}, {S{\'a}nchez Almeida}, {S{\'a}nchez}, \& {Mu{\~n}oz-Tu{\~n}{\'o}n}}]{Sanchez_2024b}
{S{\'a}nchez-Menguiano}, L., {S{\'a}nchez Almeida}, J., {S{\'a}nchez}, S.~F., \& {Mu{\~n}oz-Tu{\~n}{\'o}n}, C. 2024{\natexlab{b}}, \aap, 681, A121, \dodoi{10.1051/0004-6361/202346708}

\bibitem[{Sanders {et~al.}(2023{\natexlab{a}})Sanders, Shapley, Topping, Reddy, \& Brammer}]{Sanders_2023}
Sanders, R.~L., Shapley, A.~E., Topping, M.~W., Reddy, N.~A., \& Brammer, G.~B. 2023{\natexlab{a}}, Direct $T_e$-based Metallicities of z=2-9 Galaxies with JWST/NIRSpec: Empirical Metallicity Calibrations Applicable from Reionization to Cosmic Noon.
\newblock \doarXiv{2303.08149}

\bibitem[{Sanders {et~al.}(2023{\natexlab{b}})Sanders, Shapley, Topping, Reddy, \& Brammer}]{Sanders_2023_BPT}
---. 2023{\natexlab{b}}, Excitation and Ionization Properties of Star-forming Galaxies at z=2.0-9.3 with JWST/NIRSpec.
\newblock \doarXiv{2301.06696}

\bibitem[{{Sanders} {et~al.}(2021){Sanders}, {Shapley}, {Jones}, {Reddy}, {Kriek}, {Siana}, {Coil}, {Mobasher}, {Shivaei}, {Dav{\'e}}, {Azadi}, {Price}, {Leung}, {Freeman}, {Fetherolf}, {de Groot}, {Zick}, \& {Barro}}]{Sanders_2021}
{Sanders}, R.~L., {Shapley}, A.~E., {Jones}, T., {et~al.} 2021, \apj, 914, 19, \dodoi{10.3847/1538-4357/abf4c1}

\bibitem[{{Scannapieco} {et~al.}(2005){Scannapieco}, {Tissera}, {White}, \& {Springel}}]{Scannapieco_2005}
{Scannapieco}, C., {Tissera}, P.~B., {White}, S.~D.~M., \& {Springel}, V. 2005, \mnras, 364, 552, \dodoi{10.1111/j.1365-2966.2005.09574.x}

\bibitem[{Schaye {et~al.}(2010)Schaye, Vecchia, Booth, Wiersma, Theuns, Haas, Bertone, Duffy, McCarthy, \& van~de Voort}]{Schaye_2010}
Schaye, J., Vecchia, C.~D., Booth, C.~M., {et~al.} 2010, Monthly Notices of the Royal Astronomical Society, 402, 1536–1560, \dodoi{10.1111/j.1365-2966.2009.16029.x}

\bibitem[{Steidel {et~al.}(2014)Steidel, Rudie, Strom, Pettini, Reddy, Shapley, Trainor, Erb, Turner, Konidaris, Kulas, Mace, Matthews, \& McLean}]{Steidel_2014}
Steidel, C.~C., Rudie, G.~C., Strom, A.~L., {et~al.} 2014, The Astrophysical Journal, 795, 165, \dodoi{10.1088/0004-637x/795/2/165}

\bibitem[{Thompson \& Heckman(2024)}]{Thompson_2024}
Thompson, T.~A., \& Heckman, T.~M. 2024, Theory and Observation of Winds from Star-Forming Galaxies.
\newblock \doarXiv{2406.08561}

\bibitem[{{Tremonti} {et~al.}(2004){Tremonti}, {Heckman}, {Kauffmann}, {Brinchmann}, {Charlot}, {White}, {Seibert}, {Peng}, {Schlegel}, {Uomoto}, {Fukugita}, \& {Brinkmann}}]{Tremonti_2004}
{Tremonti}, C.~A., {Heckman}, T.~M., {Kauffmann}, G., {et~al.} 2004, \apj, 613, 898, \dodoi{10.1086/423264}

\bibitem[{{Valentino} {et~al.}(2023){Valentino}, {Brammer}, {Gould}, {Kokorev}, {Fujimoto}, {Jespersen}, {Vijayan}, {Weaver}, {Ito}, {Tanaka}, {Ilbert}, {Magdis}, {Whitaker}, {Faisst}, {Gallazzi}, {Gillman}, {Gim{\'e}nez-Arteaga}, {G{\'o}mez-Guijarro}, {Kubo}, {Heintz}, {Hirschmann}, {Oesch}, {Onodera}, {Rizzo}, {Lee}, {Strait}, \& {Toft}}]{Valentino_2023}
{Valentino}, F., {Brammer}, G., {Gould}, K. M.~L., {et~al.} 2023, \apj, 947, 20, \dodoi{10.3847/1538-4357/acbefa}

\bibitem[{{van de Voort} {et~al.}(2011){van de Voort}, {Schaye}, {Booth}, {Haas}, \& {Dalla Vecchia}}]{Voort_2011}
{van de Voort}, F., {Schaye}, J., {Booth}, C.~M., {Haas}, M.~R., \& {Dalla Vecchia}, C. 2011, \mnras, 414, 2458, \dodoi{10.1111/j.1365-2966.2011.18565.x}

\bibitem[{{Veilleux} {et~al.}(2020){Veilleux}, {Maiolino}, {Bolatto}, \& {Aalto}}]{Veilleux_2020}
{Veilleux}, S., {Maiolino}, R., {Bolatto}, A.~D., \& {Aalto}, S. 2020, \aapr, 28, 2, \dodoi{10.1007/s00159-019-0121-9}

\bibitem[{{Wang} {et~al.}(2018){Wang}, {Kong}, \& {Pan}}]{Wang-18}
{Wang}, E., {Kong}, X., \& {Pan}, Z. 2018, \apj, 865, 49, \dodoi{10.3847/1538-4357/aadb9e}

\bibitem[{{Wang} \& {Lilly}(2021)}]{Wang_2021}
{Wang}, E., \& {Lilly}, S.~J. 2021, \apj, 910, 137, \dodoi{10.3847/1538-4357/abe413}

\bibitem[{{Wang} \& {Lilly}(2022{\natexlab{a}})}]{Wang-22a}
---. 2022{\natexlab{a}}, \apj, 929, 95, \dodoi{10.3847/1538-4357/ac5e31}

\bibitem[{{Wang} \& {Lilly}(2022{\natexlab{b}})}]{Wang-22b}
---. 2022{\natexlab{b}}, \apj, 927, 217, \dodoi{10.3847/1538-4357/ac49ed}

\bibitem[{{Wang} {et~al.}(2019){Wang}, {Lilly}, {Pezzulli}, \& {Matthee}}]{Wang_2019}
{Wang}, E., {Lilly}, S.~J., {Pezzulli}, G., \& {Matthee}, J. 2019, \apj, 877, 132, \dodoi{10.3847/1538-4357/ab1c5b}

\bibitem[{Wilkins {et~al.}(2022)Wilkins, Vijayan, Lovell, Roper, Zackrisson, Irodotou, Seeyave, Kuusisto, Thomas, Caruana, \& Conselice}]{Wilkins_2022}
Wilkins, S.~M., Vijayan, A.~P., Lovell, C.~C., {et~al.} 2022, Monthly Notices of the Royal Astronomical Society, 518, 3935–3948, \dodoi{10.1093/mnras/stac3281}

\bibitem[{{Wuyts} {et~al.}(2014){Wuyts}, {Kurk}, {F{\"o}rster Schreiber}, {Genzel}, {Wisnioski}, {Bandara}, {Wuyts}, {Beifiori}, {Bender}, {Brammer}, {Burkert}, {Buschkamp}, {Carollo}, {Chan}, {Davies}, {Eisenhauer}, {Fossati}, {Kulkarni}, {Lang}, {Lilly}, {Lutz}, {Mancini}, {Mendel}, {Momcheva}, {Naab}, {Nelson}, {Renzini}, {Rosario}, {Saglia}, {Seitz}, {Sharples}, {Sternberg}, {Tacchella}, {Tacconi}, {van Dokkum}, \& {Wilman}}]{Wuyts_2014}
{Wuyts}, E., {Kurk}, J., {F{\"o}rster Schreiber}, N.~M., {et~al.} 2014, \apjl, 789, L40, \dodoi{10.1088/2041-8205/789/2/L40}

\bibitem[{{Zahid} {et~al.}(2014){Zahid}, {Dima}, {Kudritzki}, {Kewley}, {Geller}, {Hwang}, {Silverman}, \& {Kashino}}]{Zahid_2014}
{Zahid}, H.~J., {Dima}, G.~I., {Kudritzki}, R.-P., {et~al.} 2014, \apj, 791, 130, \dodoi{10.1088/0004-637X/791/2/130}

\end{thebibliography}

\end{document}